\documentclass[3p,preprint,12pt]{elsarticle}

\let\today\relax
\makeatletter
\def\ps@pprintTitle{%
    \let\@oddhead\@empty
    \let\@evenhead\@empty
    \def\@oddfoot{\footnotesize\itshape
         {Submitted preprint} \hfill\today}%
    \let\@evenfoot\@oddfoot
    }
\makeatother

    \usepackage{tgtermes}
    \usepackage{lineno}

    \usepackage{hhline} 
    \usepackage{float}
    \usepackage{multirow}
    \usepackage{graphicx,booktabs}
    \usepackage{xfrac}
    \usepackage[hidelinks]{hyperref}
    \usepackage{tabularx}
    \usepackage{wrapfig,booktabs}
    \usepackage{mwe}
    \usepackage{adjustbox}
    \usepackage{caption}
    \usepackage{subcaption}
    \usepackage{makecell}
    \usepackage{amsmath}
    \usepackage{amsmath,array}
    \usepackage[usenames, dvipsnames]{color}
    \usepackage[table]{xcolor}
    \usepackage{array}
    \usepackage{longtable}
    \usepackage{ragged2e}
    \usepackage{mathtools}
    \usepackage{setspace}
    \usepackage{tabu}

    \usepackage{MnSymbol}
    \usepackage{xcolor}
    \usepackage{algpseudocode}

    \usepackage{chngcntr}
    \newcommand{\etal}{\textit{et al}.}

    \newcommand{\noop}[1]{}

    \biboptions{numbers,sort&compress}

\def\etal{\mbox{\it et al.\ }}

\usepackage{multirow}

\begin{document}

\begin{frontmatter}
\title{An Automated Fully-Computational Framework to Construct Printability Maps for Additively Manufactured Metal Alloys}

\author{Sofia Sheikh$^{a}$, Meelad Ranaiefar$^{a}$, Pejman Honarmandi$^{a}$, Brent Vela$^{a}$, Peter Morcos$^{a}$, David Shoukr$^{b}$, Ibrahim Karaman$^{a}$, Alaa Elwany$^{a,b}$, Raymundo Arr\'{o}yave$^{a,b}$}
\address{$^a$Department of Materials Science and Engineering, Texas A\&M University, College Station, TX, USA}
\address{$^b$Department of Industrial and Systems Engineering, Texas A\&M University, College Station, Texas, USA}

\cortext[mycorrespondingauthor]{Corresponding author email:\textrm{sofiasheikh@tamu.edu}}

\begin{abstract}
In additive manufacturing, the optimal processing conditions need to be determined to fabricate porosity-free parts. For this purpose, the design space for an arbitrary alloy needs to be scoped and analyzed to identify the areas of defects for different laser power-scan speed combinations and can be visualized using a printability map. Constructing printability maps is typically a costly process due to the involvement of experiments, which restricts their application in high-throughput product design. To reduce the cost and effort of constructing printability maps, a fully computational framework is introduced in this work. The framework combines CALPHAD models and a reduced-order model to predict material properties. Then, an analytical thermal model, known as the Eagar-Tsai model, utilizes some of these materials' properties to calculate the melt pool geometry during the AM processes. In the end, printability maps are constructed using material properties, melt pool dimensions, and commonly used criteria for lack of fusion, balling, and keyholing defects. To validate the framework and its general application to laser powder-bed fusion alloys, five common additive manufacturing alloys, i.e. 316 Stainless Steel, Inconel 718, Ti-6Al-4V, AF96, and Ni-5Nb, are analyzed. Furthermore, NiTi-based alloys at three different compositions are evaluated to show the further extension of the framework to alloy systems at different compositions. The defect regions in these printability maps are validated with corresponding experimental observations to compare and benchmark the defect criteria and find the optimal criterion set with the maximum accuracy for each unique material composition. Furthermore, printability maps for NiTi that are obtained from our framework are used in conjunction with process maps resulting from a multi-model framework to guide the fabrication of defect-free additive manufactured parts with tailorable properties and performance.
\end{abstract}

\end{frontmatter}

\section{Introduction}
Additive manufacturing (AM) technologies have been proven across multiple applications including aerospace, defense, automotive, and biomedical industries~\cite{gibson2021additive, zhang2021efficient}. In the specific cases of metal AM processes, challenges related to the formation of defects (e.g., lack of fusion, balling, keyhole-induced porosity, and hot cracking) still pose barriers towards their full utilization in critical applications.

The manufacturing of porosity-free AM products requires careful assessment of the correlation between the defects and processing conditions (e.g., energy source laser power and scan speed, hatch spacing, substrate thickness, etc.) as well as materials properties (e.g., density, transformation temperatures, heat absorptivity, thermal conductivity, heat capacity, etc.) To address this challenge, printability maps ~\cite{johnson2019assessing, seede2020ultra, montgomery2015process, scime2019using, zhang2021efficient, xue2021controlling,atli2021laser} are constructed to find sub-regions within the processing space that result in no defects for a given material system. In prior works, experimental data is involved in the printability analyses through deriving defect criteria with the aid of experiments and/or parameter calibration of the contributing physical models, such as thermal models. However, the cost and time associated with acquiring high-quality experimental data limits the application of these analyses in high-throughput approaches for AM materials and process design. Therefore, a fully automated computational framework with reasonable accuracy is required to efficiently guide the AM design path in composition and process spaces.

In this work, we have implemented a computational framework that enables the construction of fully predictive printability maps for given alloy systems in an automated, high-throughput manner. We note that here we use physics-based models to account for different phenomena during AM processing. Within the proposed framework, materials properties and AM processing-related physical quantities are calculated given the chemistry of arbitrary alloys using CALPHAD-based databases as well as existing empirical and analytical equations. Some of these properties and quantities are subsequently used in a thermal model that accounts for the impact of process conditions and materials physical properties on melt pool characteristics. The melt pool dimensions and some of the physical quantities are then utilized in a set of inequality expressions (referred to as printability criteria in this work) to identify regions associated with different defect inducing regimes. As a case study, the implemented framework is applied on commonly fabricated AM alloys: 316 Stainless Steel, Ti-6Al-4V, Inconel 718, Ni-5Nb and AF96. Printability maps were constructed for each alloy and validated using experimental data. Furthermore, we apply the framework to a single type of alloy at different compositions to show the effectiveness of the framework to predict printability maps within alloy families. For the case study,  NiTi-based Shape Memory Alloys (SMAs) with attractive functional properties are used to construct their printability maps and validate them with available limited experimental data. The construction of the printability maps for NiTi SMAs is of high value since AM processes are among the few methods that can be used to fabricate parts with complex geometries due to the poor machinability of these alloys~\cite{zhang2021efficient, xue2021controlling, elahinia2016fabrication,xue2022laser}. We hope this case study will demonstrate the predictive capabilities of our automated computational framework as it opens a new avenues for accelerated high-throughput materials and process design in AM.

\section{Prior Work}\label{sec:prior}
Printability is understood as the characteristics of an alloy (in the case of metal AM) associated with its processability under AM conditions. A printability map is a graphical depiction of the process space. In this depiction, the different process conditions under which different types of (fabrication) defects are prevalent~\cite{johnson2019assessing} are defined by color and/or bounding-lines. In practice, a large fraction (the vast majority in most cases) of the process space tends to result in fabrication defects, manifested as porosity, which ultimately affect the integrity of the fabricated part. The identification of suitable regions for porosity-free parts is thus essential to develop robust protocols for AM experimentation. 

Prior work by the present authors~\cite{seede2020ultra} and others indeed indicate that variability in performance can be dramatically reduced when the alloys are fabricated within the printable region in process space. Recently, significant effort has been invested in developing frameworks to assess the printability of a given alloy. The present authors, for example, recently developed a framework to rapidly assess the printability region of arbitrary alloys fabricated with laser powder bed fusion AM (L-PBF) by employing physics-based thermal models for the melt pool characteristics calibrated against experiments~\cite{zhang2021efficient}. Islam \etal ~\cite{islam2022high} recently presented a similar framework in which they combined analytical models and high-throughput sample fabrication and characterization to determine the processing parameters leading to minimal fabrication defects. Rather than focusing on developing the framework in the context of a printability map, Islam and collaborators designed their experimental protocol in terms of a dimensionless quantity $\Pi=\frac{C_p P}{k v^2 h }$---where $C_p$ is the specific heat, $P$ is laser power, $k$ is thermal conductivity, $v$ is laser scan speed, and $h$ is hatch spacing---connecting the heat deposited and dissipated in the powder bed. $\Pi$ was shown to be correlated to the final density in the fabricated part.

In addition to experiment-focused frameworks, other groups have employed machine learning (ML) approaches. Du \etal recently demonstrated the use of ML approaches, combined with physics-based models and experimental data to predict the onset of balling in L-PBF metal AM~\cite{du2021physics}. With their approach, they were able to identify numerous materials and process-related variables leading to the onset of balling. Those parameters were related to a number of physical phenomena associated with melt pool instabilities. Similarly, Vela \etal developed a physics-based indicator for balling and demonstrated the predictive ability of this indicator by using it as a feature in ML-models for balling ~\cite{vela2022evaluating}. However, both Du \etal and Vela \etal only focused on the onset of composition-based balling and neither discuss other porosity- type defects such as lack of fusion and keyholing. Zhu \etal~\cite{zhu2021predictive} combined dimensional analysis with ML and experiments to derive a dimensionless quantity associated with the formation of keyhole porosity. Similar to the work by Islam \etal~\cite{islam2022high}, the dimensionless number was associated with the balance between the heat deposited into and dissipated out of the powder bed. However, similarly, Zhu \etal validated their framework only for the onset of keyholing. Very recently, Akbari \etal ~\cite{akbari2022meltpoolnet}, presented MeltpoolNet, a neural network (NN) trained against experimental data used to predict the printability map of metal alloys. While the trained models exhibited adequate \emph{average} performance, the predictions of the different characteristic regions in a printability map (e.g. lack of fusion, keyholing, balling, etc.) exhibited significant pathologies, perhaps due to the sparsity of the training dataset. Furthermore, the printability maps predicted showed insufficient predictability compared to experimentally-derived printability maps in literature, which can once again be attributed to the sparsity of training data. Moreover, the NN models lacked in interpretability. Comparably, Scime \etal~\cite{scime2019using} used feature extraction methods and an unsupervised ML approach to predict keyholing and balling formation in the process space for Inconel 718. The framework we present suggests criteria for lack of fusion, keyholing and balling, in terms of processing parameters and material properties.

While the previous approaches constitute significant progress towards the assessment of printability, it is of interest to examine the extent to which an entirely physics-based approach, in which physics-based thermal models, materials properties and physics-derived criteria can be used to predict printability  maps of arbitrary alloy chemistries. We note that Zhu \etal~\cite{zhu2021predictive} recently presented a work in the same spirit as the present contribution, applied to NiTi SMAs. Similarly, Johnson \etal~\cite{johnson2019assessing} and Zhang \etal~\cite{zhang2021efficient} have shown that printability maps can be derived from using a FEM method and/or a simple Eagar-Tsai model that are calibrated again experiments. However, in these works only one set of criteria for defects (based on the melt pool geometry and processing parameters) is used to evaluate the printability map. Furthermore, the authors used an extremely limited amount of experimental data for validation. Here, we deploy physics-based models to investigate the process maps for a variety of alloys. We benchmark different sets of criteria for the onset of fabrication defects (total of 12) and compare the resulting maps with an exhaustive validation dataset.

\section{General Framework}\label{sec:framework}

\begin{figure*}[h]
    \centering
    \includegraphics[width=0.85\textwidth]{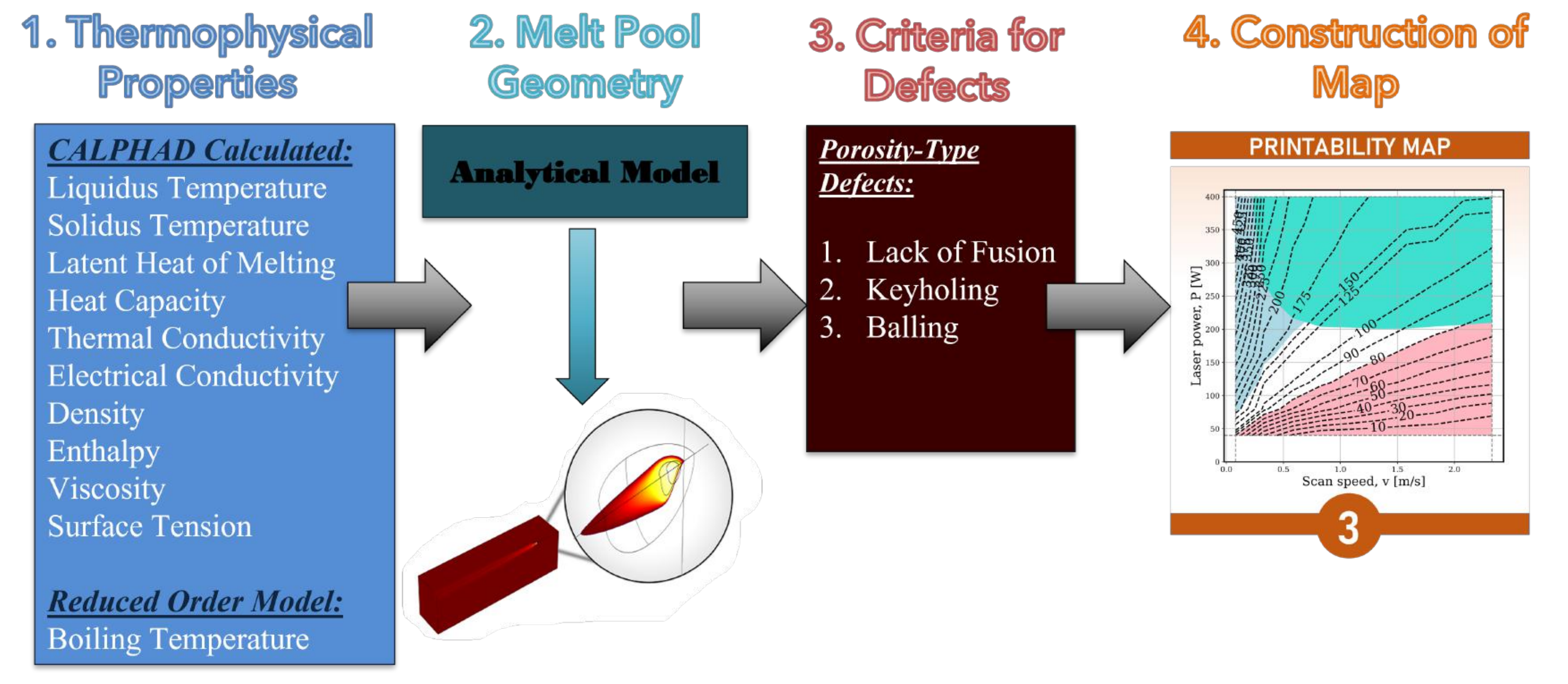}
    \caption{The framework introduced can be separated into four different stages: calculating thermophysical properties for the material system using CALPHAD-based approach or reduced order models, calculating the melt pool geometry, selecting criteria for the defects and constructing printability maps.}
    \label{fig:steps}
\end{figure*}

The general framework for the construction of a computational printability map consists of four important stages: calculating thermophysical properties, calculating the melt pool dimensions, selecting criteria for defects, and constructing the printability maps as shown in Figure \ref{fig:steps}. The framework requires the input of chemistry and process conditions (such as the ranges of laser power and scan speed to be considered in the design space). Once the chemistry and process window has been fully defined, the framework proceeds to predict the printability maps accounting for different criteria for the onset of (macroscopic) printing defects. Each stage of the framework is discussed in further detail below.

\subsection{Calculation of Thermophysical Properties and Melt Pool Geometry}
To calculate the thermophysical properties relevant to AM, several models and formulations were used. Our framework uses Thermo-Calc to derive CALPHAD-based thermophysical properties using the \emph{Property}, \emph{Equilibrium} and \emph{Scheil} models. The models in turn use three different thermodynamic databases depending on the alloy system in question: TCHEA5, TCAL7 and TCNI11. For the case studies presented, the Thermo-Calc calculations are simplified where only two Thermo-Calc models and a single database are used: Property and Equilibrium Models with the TCHEA5 thermodynamic database. A variety of material properties relevant to the AM process were calculated, such as the liquidus and solidus temperatures, latent heat of melting, specific heat, etc. At the liquidus, solidus, and room temperatures, the following CALPHAD-derived properties were calculated: electric conductivity, heat capacity, thermal conductivity, thermal diffusivity, thermal resistivity, density, enthalpy, dynamic and kinetic viscosity, and surface tension. These properties, as will be discussed below, were in turn used as either input to the thermal models or to estimate physical quantities or dimensionless numbers related to AM. Other than CALPHAD-based material properties, a rule-of-mixtures model (ROM), $P_{alloy} = \Sigma_{i}^n x_{i}P_{i}$, where $i$ is the set of atomic species, $x$ is the composition value and $P$ is the material property, is used to calculate the (average, representative) melting temperature, boiling temperature, density and molecular weight.

Using the CALPHAD-based material properties and ROM calculated properties, additional attributes were calculated such as the solidification range which can be calculated using the formulation, $T_{liquidus} - T_{solidus}$. The total enthalpy can be defined as $H_{liquidus} - H_{RT}$, where $H_{liquidus}$ is the enthalpy at liquidus and $H_{RT}$ is the enthalpy at room temperature. With the total enthalpy, the effective heat capacity can be calculated using $H_{total}/(T_{liquidus} - T_{RT})$ where T$_{RT}$ is 298 K. The melting enthalpy and boiling enthalpy were quantified using $H_{liquidus} - H_{solidus}$, and $10 \times R \times T_{liquidus}$, respectively. H$_{solidus}$ is the enthalpy at the solidus temperature and R is the gas constant. Furthermore, the enthalpy at boiling and enthalpy after boiling was calculated using $H_{liquidus} + C_{p,liquidus} \times (T_{boiling} - T_{liquidus})$ and $H_{at-boiling} + H_{boiling}$, where C$_{p,liquidus}$ is the specific heat at the liquidus temperature, $T_{boiling}$ is the boiling temperature, $H_{at-boiling}$ is the enthalpy at boiling and $H_{boiling}$ is the boiling enthalpy. 

\subsection{Calculating the Melt Pool Geometry}
Using the material properties derived from CALPHAD and ROM models, melt pool dimensions for alloys processed using L-PBF were calculated using the Eagar–Tsai model (E-T)~\cite{eagar1983temperature}. The E-T model was originally used to (approximately) predict the quasi-steady-state temperature fields during (some) welding processes. Recently, it has been widely applied to predict melt pool characteristics during metal AM processes. The E-T model predicts the shape of the isotherms around a moving heat source in a semi-infinite medium as a function of both thermo-physical properties and processing conditions, including laser power [W], scan speed [m/s] and beam diameter [$\mu$m]. The material inputs for the E- T analytical model are thermal conductivity [W/m K], density [kg/m$^{3}$], specific heat [J/kg K], melting (liquidus) temperature [K], and laser absorptivity. The material properties, except absorptivity, were calculated at the liquidus temperature using ThermoCalc. For the specific heat, we considered the effective specific heat, as this is the most accurate measure of the ability of a material to absorb heat through the entire heating and melting process during AM. The laser absorptivity, A, was estimated according to Drude’s theory using Equation \ref{absorp} where $\rho_0$ is the electrical resisitivity [$\Omega$ m] and $\lambda$ is the laser wavelength [m]~\cite{schuocker1998handbook,letenneur2019optimization,zhu2021predictive}.

					\begin{equation}\label{absorp}
					A = 0.365\sqrt{\frac{\rho_{0}}{\lambda}}
					\end{equation}

The E-T analytical model solves the heat conduction equation by considering a Gaussian-distributed heat source that moves over a semi–infinite plate and predicts the temperature distribution across the plate. Given the resulting temperature distribution and the material's melting temperature, the melt pool dimensions---length, width, and depth---can be calculated. Key assumptions of the E-T model include the neglecting of temperature-dependent thermal properties, the latent heats of melting and evaporation, and the assumption of a quasi-steady state. 

A further limitation of the E-T model is that when the melt pool changes from conduction mode to keyhole mode, the model underestimates the melt pool depth due to the missing physics associated with the keyholing mode (i.e., further deepening of the melt pool due to recoil pressure resulting from metal evaporation) in this model~\cite{zhu2021predictive}. 

To account for this keyhole mode, the melt pool depth is estimated using a model proposed by Gladush and Smurov (G-S model) instead~\cite{gladush2011properties}. Similar to the E-T model, the G-S model was also derived for the welding process and has been shown to be applicable to L-PBF. The G-S model is based on the thermal balance and mechanical equilibrium of a gas–vapor keyhole. Zhu \etal~\cite{zhu2021predictive} and Honarmandi \etal~\cite{honarmandi2021rigorous} have recently shown that the corrected depth using the G-S model yields a more accurate prediction of the melt pool depth after keyholing in comparison to the depth predicted by the E-T model. 

The G-S melt pool depth, $D$, is calculated in terms of processing and material properties as shown in Equation \ref{G-S depth} where $P$ is the laser power [W], $k$ is the thermal conductivity [W/m K], $a_{1}$ is the beam size [m] and $\alpha$ corresponds to the thermal diffusivity [m$^2$/s]. Using the corrected depth from Equation \ref{G-S depth}, the criteria for balling and lack of fusion are then evaluated using the corrected depth to define regions in which different (macroscopic) printing defects are likely to be prevalent.

\begin{equation}\label{G-S depth}
    D = \frac{AP}{2\pi kT_{boiling}}\ln(\frac{a_{1} + \frac{\alpha}{v}}{a_{1}})
\end{equation}

\subsection{Construction of Printability Maps}
After predicting the melt pool profile, various criteria for keyholing, lack of fusion, and balling are evaluated, as mentioned above. The union of these criteria is used to define regions in the processing space where the defect occurs and is projected on a 2D dimensional space to construct a printability map. However, over the past few years, multiple criteria for the onset of the aforementioned defects have been used in the literature. One of the main objectives of the present work is to benchmark the different sets of criteria by comparing their predictions with the available experimental data. 

Each set of criteria was evaluated by considering the prediction of the onset of a given defect as the (positive) outcome of a binary classifier separating the processing space on whether a point in $P-v$ coordinates belonged to a given defect class. The combination of physics-based models and criteria were evaluated in terms of the trustworthiness of the model when predicting the onset of a manufacturing defect (\emph{precision}), its ability to map the entire region corresponding to a specific defect (\emph{recall}), as well as its overall \emph{accuracy}.

For lack of fusion, two criteria were evaluated (Equation \ref{lof1} and Equation \ref{lof2}), where $D$ is the melt pool depth, $W$ is the melt pool width and $t$ is the powder layer thickness: 
					\begin{equation}\label{lof1}
					    D \leq t
					\end{equation}
					
					\begin{equation}\label{lof2}
					    (\frac{h}{W})^{2} + \frac{t}{t+D} \geq 1
					\end{equation}
					
Lack of fusion occurs when insufficient energy is deposited onto the material, which results in an incomplete bonding with the underlying layers. Therefore, if the melt pool depth is less than the powder layer thickness, the melt pool will not fully bond to the substrate and/or previous layer leading to a lack of fusion defect establishing Equation \ref{lof1} as one of the criterion~\cite{zhang2021efficient}. Equation \ref{lof2} defines the second criterion for lack of fusion introduced by Zhu \etal~\cite{zhu2021predictive}. In this case, it is stated that lack of fusion may occur when the hatch spacing, $h$, is greater than the maximum hatch spacing, $h_{max}$~\cite{zhu2021predictive,seede2020ultra} beyond which porosity due to insufficient overlapping melt tracks occurs. To avoid a lack of fusion-induced porosity, the hatch spacing needs to provide a good join between adjacent tracks, and if not done so, it would lead to incomplete bonding  between tracks and prior layers. The maximum hatch spacing can be calculated using Equation \ref{hmax}~\cite{seede2020ultra}.
					\begin{equation}\label{hmax}
					    h_{max}  = W\sqrt{1-\frac{t}{t+D}}
					\end{equation}
Using Equation \ref{hmax}, it is possible to estimate the maximum hatch spacing that would still result in fully dense layers at different locations of the process map.

Balling occurs when the melt pool breaks into droplets instead of a continuous pool due to Plateau-–-Rayleigh capillary instability. The instability is observed at high laser power and scan speeds and can be evaluated using a criterion by a ratio of the melt pool length to width ($L/W$). The threshold value for the balling criterion in literature is defined to be typical  $2 \leq L/W \leq \pi$. Based on a comparison of experiments for the general AM alloys, the threshold value of 2.3 was determined. Furthermore, the threshold value at 2.3 was proposed as a good estimate by Zhang \etal~\cite{zhang2021efficient} and Johnson \etal~\cite{johnson2019assessing} as  shown in Equation \ref{ball1}. Yadroisev \etal~\cite{yadroitsev2010single} showed that for single tracks, the threshold value between the stability zone and the instability zones of a melt pool could be defined using Equation \ref{ball2}, which is the second criterion we took into consideration for balling.

					\begin{equation}\label{ball1}
					    \frac{L}{W} \geq 2.3
					\end{equation}
                    \begin{equation}\label{ball2}
                        \frac{\pi W}{L} < \sqrt{\frac{2}{3}}
                    \end{equation}
                    
Keyholing-induced pores occur when vapor cavities form (and become trapped) due to rapid evaporation of the molten liquid that causes the deep penetration of the molten material by the recoil pressure induced on the melt pool surface. Therefore, the resultant melt pool depth is larger in comparison to the depth during the conduction mode. Ultimately, the collapse of the formed cavities results in voids. This keyholing mode is located in the high laser power, low scan speed region in the processing space.

Johnson \etal~\cite{johnson2019assessing} used a criterion for keyholing based on the ratio between the depth and width of the melt pool. Based on the comparison of experiments, they determined a threshold value of 1.5 for keyholing from empirical observations and relatively simple geometrical considerations. Using the same theory with the existing experimental data points for the general AM alloys, a threshold value of 2.75 for keyholing was set as shown in Equation \ref{key1-Gen}. However, for NiTi-based alloys, Zhang \etal~\cite{zhang2021efficient} revised the threshold value to 2 as shown in Equation \ref{key1} based on experimental measurements for NiTi alloy system, which is the criterion we considered for keyholing in our case study for NiTi. For the purposes of this study, the threshold values were kept constant at 2.0 and 2.75 for the general AM alloys and NiTi. However, for arbitrary compositions, a variety of keyhole threshold values can be analyzed to understand the printability region of the material. The printability maps are a design tool that is expected to be used as a tool to gain insight into the processing space of a material under AM conditions. 

\begin{equation}\label{key1-Gen}
\frac{W}{D} \leq 2.75
\end{equation}

\begin{equation}\label{key1}
\frac{W}{D} \leq 2.0
\end{equation}

There are other criteria, beyond the geometry of the melt pool, that consider material properties and processing parameters that can be used to estimate the onset of keyholing. King \etal~\cite{king2014observation,rubenchik2018scaling}, for example, showed a positive correlation between normalized enthalpy and the melt pool depth, and thus, to keyholing. Using the relationship, they derived a criterion for keyholing, as shown in Equation \ref{key2}. This criterion utilizes the specific enthalpy, $h_s = \rho C_p T_{liquidus}$ ($\rho$ is the density [kg/m$^3$], C$_p$ is the effective specific heat [J/kg K] and T$_{liquidus}$ is the liquidus temperature). Here, we note that the heat capacity used by King \etal, in the original formulation considers the heat capacity for the solid phase. This underestimates the ability of a metal to absorb energy from the laser and here we thus replaced this heat capacity with the \emph{effective heat capacity} accounting for the sensible and latent heat of an alloy from room temperature until the liquidus.
 
\begin{equation}\label{key2}
   \frac{\Delta H}{h_s} = \frac{AP}{\pi h_s \sqrt{\alpha va^3}} > \frac{\pi T_{boiling}}{T_{liquidus}}
\end{equation}
 
Recently, using dimensional analysis and a modification of the Buckingham-Pi theorem, Gan \etal~\cite{gan2021universal} also derived a universal keyholing criterion capable of defining, with a single metric, regions of conduction, transition and keyhole mode of the molten pool. The dimensionless keyhole criterion, $Ke$, as defined in Equation \ref{Ke}, is determined by processing parameters, i.e., the laser power ($P$ [W]), scan speed ($v$ [m/s]) and beam radius ($r_0$ [m]) as well as material properties, i.e., the absorptivity ($\eta$),liquidus temperature ($T_{liquidus}$ [K]), substrate temperature ($T_0$ [K]), density ($\rho$ [kg/m$^3$]), heat capacity ($C_p$ [J/kgK]), and thermal diffusivity ($\alpha$ [m$^2$/s]). In their work, using a limited experimental dataset, Gan \etal arrived at $Ke > 6.0$ as the threshold for keyhole mode in L-PBF processes.

\begin{equation}\label{Ke}
    Ke = \frac{\eta P}{(T_{liquidus} - T_0) \pi \rho C_p \sqrt{\alpha v r^3_{0}}}>6
\end{equation}

In total, in this work, we will consider two criteria for each lack of fusion and balling, and three criteria for keyholing. A total of 12 printability maps were constructed from the different combinations of these defect criteria. Additionally, for each alloy system we enriched the predicted printability maps with contour lines that define maximum allowed hatch spacing to prevent porosity due to the lack of overlap of melt pool tracks, as defined by Equation \ref{hmax}. Table \ref{table:labels} summarizes the different criteria used.

\renewcommand{\arraystretch}{1.5}
\begin{table*}
    \centering
    \setlength\extrarowheight{5pt}
    \begin{tabularx}{0.9\textwidth} { 
      | >{\centering\arraybackslash}X 
      | >{\centering\arraybackslash}X 
      | >{\centering\arraybackslash}X | }
     \hline
     \textbf{Defect Type} & \textbf{Label} & \textbf{Equation} \\
    \hline
    Lack of Fusion & LOF1 &     D $\leq$ t \\
    & LOF2 &  ($\frac{h}{W})^{2}$ + $\frac{t}{t+D} \geq$ 1 \\
    \hline
    Balling & Ball1 &  $\frac{L}{W}$ $\geq$ 2.3 \\
    & Ball2 & $\frac{\pi W}{L} < \sqrt{\frac{2}{3}}$ \\
    \hline
    Keyholing  & KH1  & $\frac{W}{D}$ $\leq$ 2.5 (for general AM alloys)\\
    && $\frac{W}{D}$ $\leq$ 2.0 (for NiTi-based alloys)  \\ 
    & KH2 & $\frac{\Delta H}{h_s} = \frac{AP}{\pi h_s \sqrt{\alpha va^3}} > \frac{\pi T_{boiling}}{T_{liquidus}}$ \\
    & KH3 & $Ke = \frac{\eta P}{(T_{liquidus} - T_0) \pi \rho C_p \sqrt{\alpha v r^3_{0}}}>6$ \\
    \hline
    \end{tabularx}
    \caption{The equation and labels for each respective defect used to graph the printability maps in Figure \ref{fig:50.3 maps}, Figure \ref{fig:50.8 maps}, and Figure \ref{fig:51.2 maps}.}
    \label{table:labels}
\end{table*}

Based on the overall results of the case study, the optimal combination of criteria for lack of fusion, keyholing, and balling was determined. In this regard, for each printability map that is constructed, the experimental points derived from the in-house experiments are overlaid in order to determine the validity of the map by calculating different statistical metrics related to the ability of the combination of physics-based models and criteria to correctly label the (printing) outcome of different locations within the printability map.

\subsection{Evaluating the Performance of Proposed Printability Maps}

In order to study the performance of printability maps constructed by our proposed framework, we converted the problem into binary classification problems where each label in printability maps either exists or not, e.g., lack of fusion/not lack of fusion. The points corresponding to the label that is being examined were taken as the positive class and the remaining points were considered as the negative (complementary) class/label. There are several classification performance metrics that can be used to evaluate these classification problems, including \emph{precision} (Equation \ref{eq:Pre}), \emph{recall} (Equation \ref{eq:Rec}), and \emph{accuracy} (Equation \ref{eq:Acc}).

\begin{equation} \label{eq:Pre}
    Precision = \frac{TP}{(TP+FP)}
\end{equation}
\begin{equation} \label{eq:Rec}
    Recall = \frac{TP}{(TP + FN)}
\end{equation}
\begin{equation}
    Accuracy = \frac{TP + TN}{TP + FP + TN + FN}
    \label{eq:Acc}
\end{equation}
where P and N are the number of points in the positive and negative classes, and TP, TN, FP, and FN denote the number of positive points that are correctly classified, negative points that are correctly classified, negative points that are misclassified, and positive points that are misclassified, respectively.

Once a model-criteria combination predicts that a particular region in the processing space leads to a defect (or no defect), \emph{precision} measures the ability of the model to predict the positive class. 100\% precision would mean that every single point that is predicted to belong to a defect class indeed belongs to that defect class. \emph{Recall}, on the other hand, measures the ability of the model-criteria combination to capture the entirety of the experimentally determined region in the processing space where the defect is prevalent or in other words the defect region being assessed. In this case, 100\% recall means that the model-criteria combination was able to correctly map the onset of a given defect. The \emph{precision} and \emph{recall} metrics are complemented by the overall \emph{accuracy} of the prediction.

For each printability map constructed, the values of the above metrics obtained from four binary classification cases are averaged to find the average performance of each criteria combination. Note that the averaging in this case was carried out by considering the prevalence of a particular type of defect within the experimental data set and in the rest of the paper, the weighted average will be referred as average. Using the data points from our internal L-PBF database for the respective alloy systems, the maximum average accuracy value was considered to find the optimal combination of criteria in constructing the printability maps. However, the recall and precision are still presented to fully assess the printability map performance. 

\section{Case Study with General AM Alloys}
\label{sec: General AM}

\subsection{Determining Process Parameters}

To show the generality and effectiveness of the framework to L-PBF AM processed alloys, the proposed computational framework to predict printability maps was applied to common AM alloys: 316 Stainless Steel, Ti-6Al-4V, Inconel 718, Ni-5Nb and AF96. For these alloys, a 20 by 15  design grid of experiments was utilized for laser power and scan speed with a constant beam diameter, powder layer thickness and hatch spacing at 80 $\mu$m, 30 $\mu$m and 70 $\mu$m, respectively.  

\begin{figure*}
    \centering
    \includegraphics[width=0.85\textwidth]{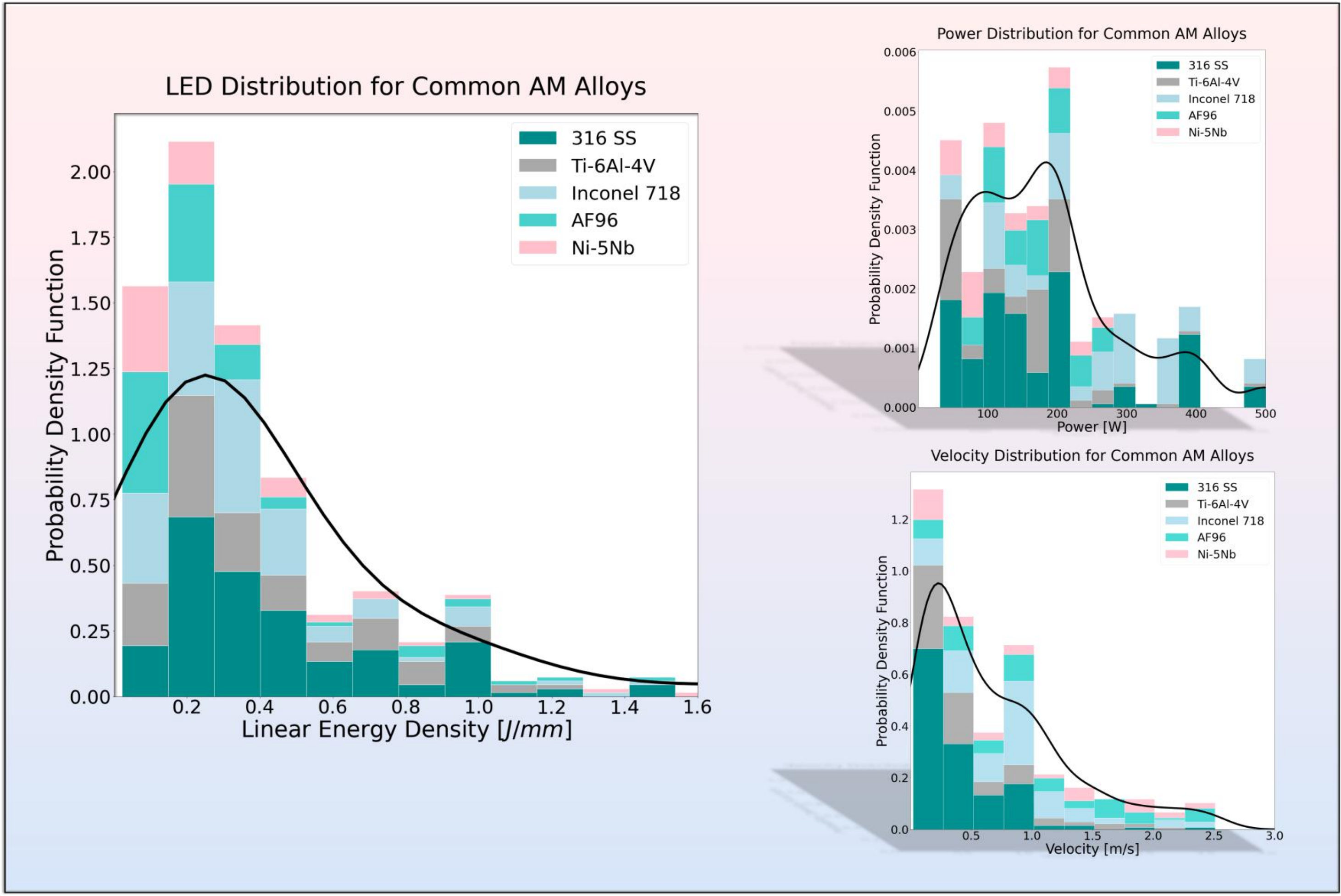}
    \caption{The histogram and kernel density estimate (KDE) plots for laser power, scan speed, and linear energy density (LED) show the distribution of values used to print for a variety of alloys, as reported in the literature. The density represents the probability density function of the parameter on the x-axis in the plot.}
    \label{fig:database_general}
\end{figure*}

To narrow down the processing space to construct the printability maps for any arbitrary alloy, experimental data collected from a literature survey as well as in-house experiments at Texas A\&M University for L-PBF AM processes has been captured in a database. Each point in the chemistry-processing space is represented by close to 500 features related to different important factors in metal AM, ranging from conditions of the feedstock, materials properties, process conditions, printing outcome, (measured) properties and performance of the printed part, etc. The distribution of the number of single-track data points extracted for each alloy is shown in Table \ref{table:exp_dis}. The experimental points for 316 stainless steel ~\cite{king2014observation,liu2015investigation,zhang2013study,philo2019pragmatic,liang2021prediction,shi2020properties,antony2014numerical,yadroitsev2013energy,steinfeld2015role,bertoli2017limitations}, Inconel 718~\cite{yang2018printability,onuike2019additive,sadowski2016optimizing,scime2019melt,xia2017porosity,yang2019controllable,zheng2021role} and Ti-6Al-4V ~\cite{karayagiz2019numerical,yang2016role,parry2016understanding,zheng2019melt,mahmoudi2018multivariate,he2019melt} originates from published literature, while points for Ni-5Nb (weight \%)~\cite{johnson2019assessing,karayagiz2020finite,atli2021laser} and AF96 ~\cite{seede2020ultra} originates from both in-house experiments as well as experiments from literature. Figure \ref{fig:database_general} displays the range of processing parameter values used in the data set for validation. Based on Figure \ref{fig:database_general}, the ranges for processing parameters considered were 40-400 W and 0.05 – 3.00 m/s for laser power and scan speed, respectively.  

\renewcommand{\arraystretch}{1.5}
\begin{table*}
    \centering
    \setlength\extrarowheight{5pt}
    \begin{tabularx}{0.9\textwidth} { 
      | >{\centering\arraybackslash}X 
      | >{\centering\arraybackslash}X | }
     \hline
     \textbf{Alloy Type} & \textbf{Single Track Datasets}\\
    \hline
    316 Stainless Steel & 83 \\
    \hline
    Inconel 718 & 86 \\
    \hline
    Ti-6Al-4V  & 41 \\
    \hline
    Ni-5Nb & 49\\
    \hline
    AF96 & 64\\
    \hline
    \end{tabularx}
    \caption{The table displays the number of experimental points used for each alloy to validate the printability maps}. 
    \label{table:exp_dis}
\end{table*}

 Figure \ref{fig:database_general} also shows that LED (defined as $LED=P/v$, with $P$ being laser power and $v$ being scan speed) peaks between 0.2 - 0.4 $\frac{J}{mm}$ and decays rather quickly after exceeding $\sim~0.6~\frac{J}{mm}$. In contrast, the distribution of laser power used in the materials was more evenly distributed, ranging from 50 to $\sim$250 W. In the case of scan speed, the conditions used tended to be slightly clustered around relatively low values ($<0.5~m/s$), although there are considerable points in the processing space corresponding to high scan speeds between $1-2.5~m/s$.  After determining the processing space for a general AM alloy, the framework was used to print printability maps for the alloys. We would like to note here, that even though we are looking at single tracks, the hatch spacing is a pre-determined, constant number that can be adjusted based on analysis by the experimenter, such as can powder layer thickness. The effects of the processing parameters are analyzed in Section \label{sec: hatch-layer}.
 
\subsection{Constructing and Evaluating Printability Maps}
\begin{figure*}
    \centering
    \includegraphics[width=0.75\textwidth]{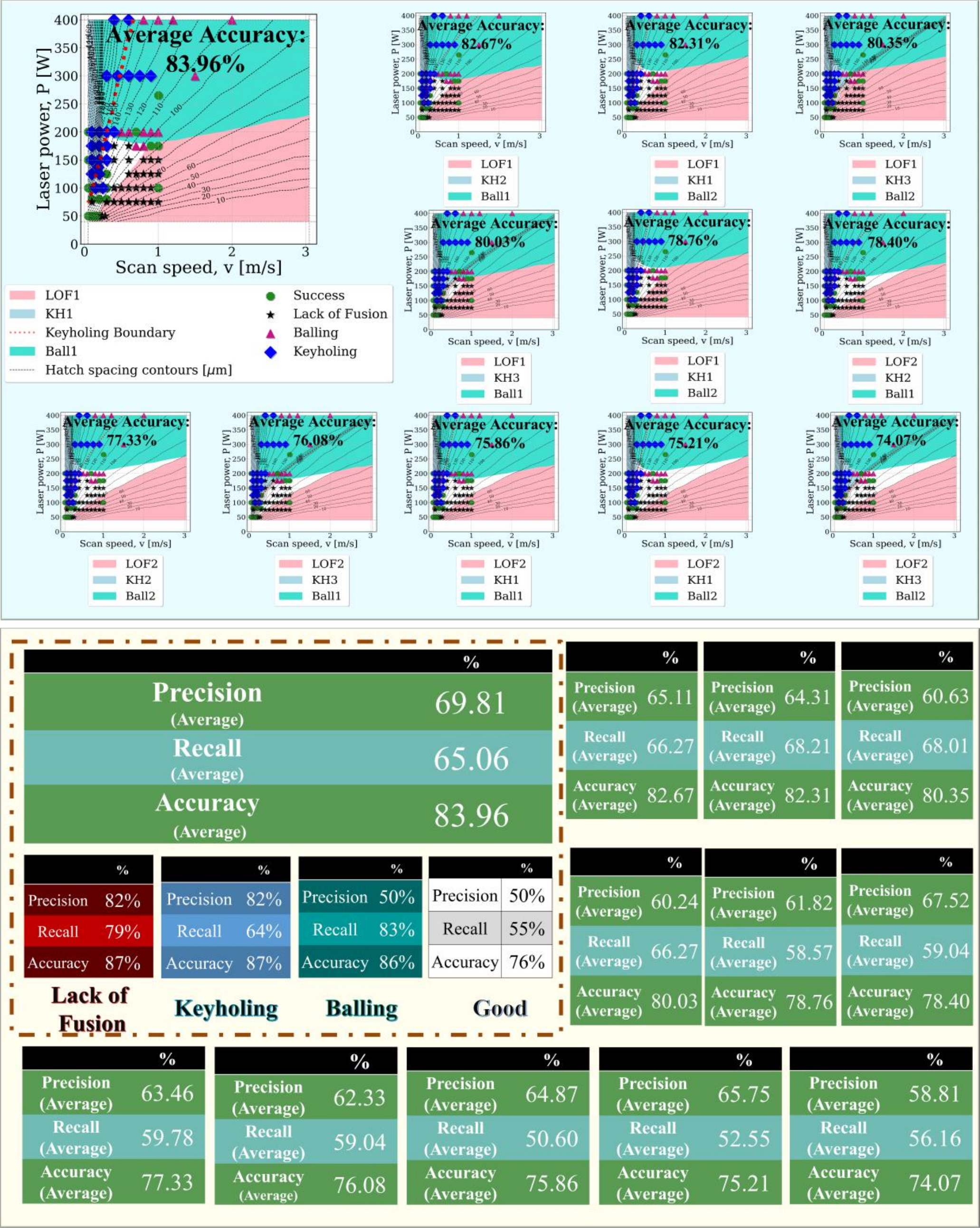}
    \caption{The printability maps produced using different combinations of defect criteria for lack of fusion, balling, and keyholing for 316 Stainless Steel.Based on the experimentally observed data overlaid on the printability maps and the package prediction in each case, the precision, recall, and accuracy were calculated for four binary classification problems defined based on the presence and absence of each label (i.e., lack of fusion, keyholing, balling, and success or defect-free) in the printability maps and then their averages were listed.  Each map corresponds to the different criteria combination and the maps are placed in order of highest average accuracy to lowest. The values of the precision and recall alongside the accuracy for each map are also displayed. The maximum value for average accuracy was 83.96 \%. The defect labels (i.e., LOF, Ball, and KH) are defined in Table \ref{table:labels}.}
    \label{fig:316 maps}
\end{figure*}

With the processing space scoped, the printability maps were constructed. The printability map of 316 stainless steel (316 SS) is shown in Figure \ref{fig:316 maps} where the maximum accuracy value was 83.96\%. The minimum accuracy value was 74.07\%. The combination of the criterion that gave the maximum accuracy map for 316 SS was Equation \ref{lof1} for lack of fusion, Equation \ref{key1-Gen} for keyholing and Equation \ref{ball1}. It is important to note two items here. First, the single-track experiments were all obtained from published literature, therefore there is a distribution associated with the machine and powder parameters that may have been used for the various points and account for some of the error. However, we would like to note that these printability maps are a tool that helps gain insight into how printable material is under AM conditions. Considering such, the data set for 316 SS was expansive enough that the prediction accuracy was fairly high. 

\begin{figure*}
    \centering
    \includegraphics[scale=0.65]{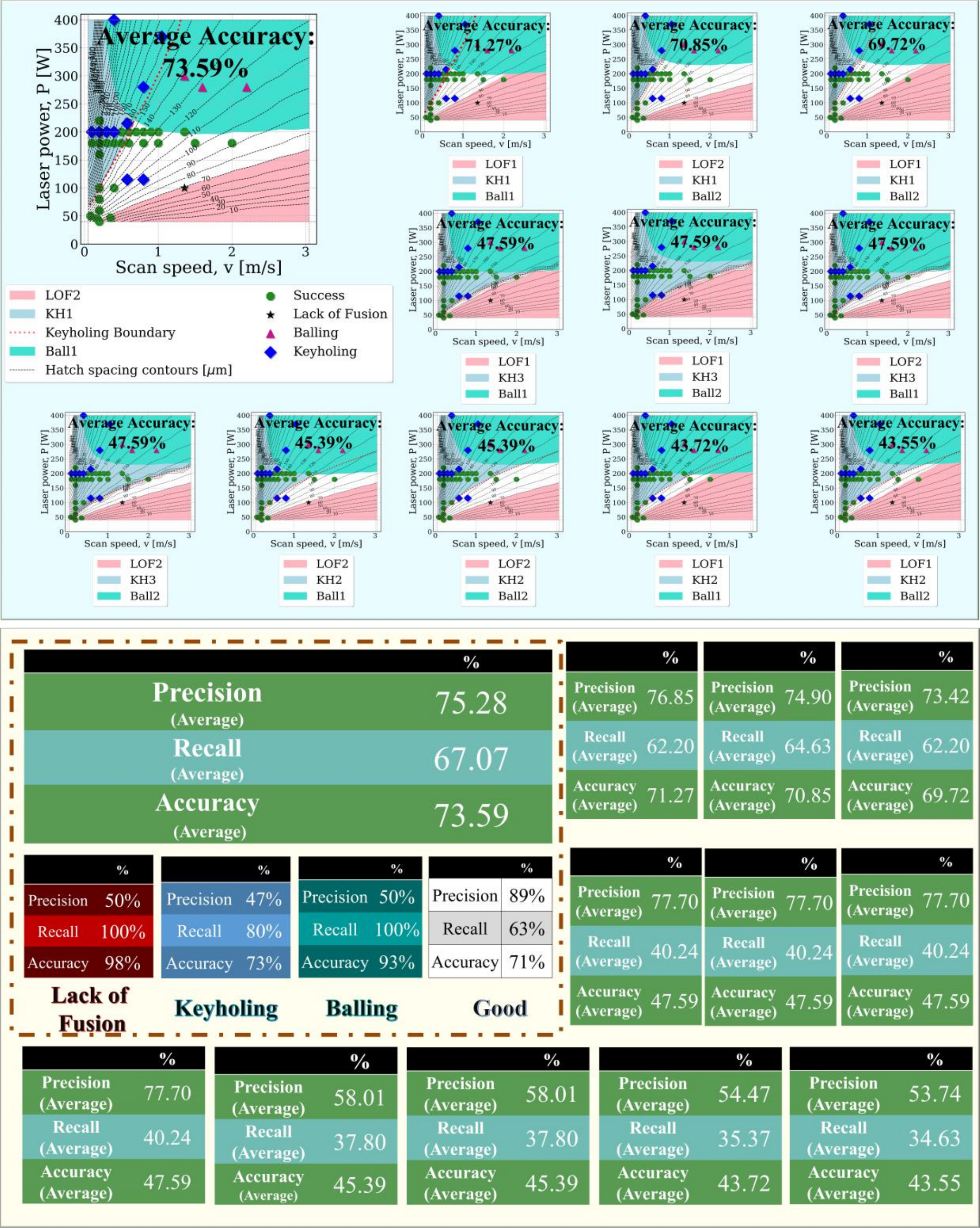}
    \caption{The printability maps produced using different combinations of defect criteria for lack of fusion, balling, and keyholing for Ti-6Al-4V.Based on the experimentally observed data and the package prediction in each case, the precision, recall, and accuracy were calculated for four binary classification problems defined based on the presence and absence of each label (i.e., lack of fusion, keyholing, balling, and success or defect-free) in the printability maps and then their averages were listed. Each map corresponds to the different criteria combination and the maps are placed in order of highest average accuracy to lowest. The values of the precision and recall alongside the accuracy for each map are also displayed. The values of the precision and recall for each map are also displayed. The maximum value for average accuracy was 73.59 \%. The defect labels (i.e., LOF, Ball, and KH) are defined in Table \ref{table:labels}.}
    \label{fig:Ti64 maps}
\end{figure*}

The printability map of Ti-6Al-4V as shown in Figure \ref{fig:Ti64 maps} showed a maximum accuracy of 73.59 \%. It is important to note that majority of the available data published for Ti-6Al-4V report successful prints rather than prints with defect, especially the onset of lack of fusion is under sampled. Therefore, the accuracy can essentially be increased with the collection of more thorough data. However, given the different AM fabrication parameters as well as powder parameters of published data, 73.59 \% is a reasonable accuracy. The defect criteria set combination that gave the maximum accuracy was Equation \ref{lof2}, Equation \ref{key1-Gen} and \ref{ball1} for lack of fusion, keyholing and balling, respectively.

\begin{figure*}
    \centering
    \includegraphics[scale=0.65]{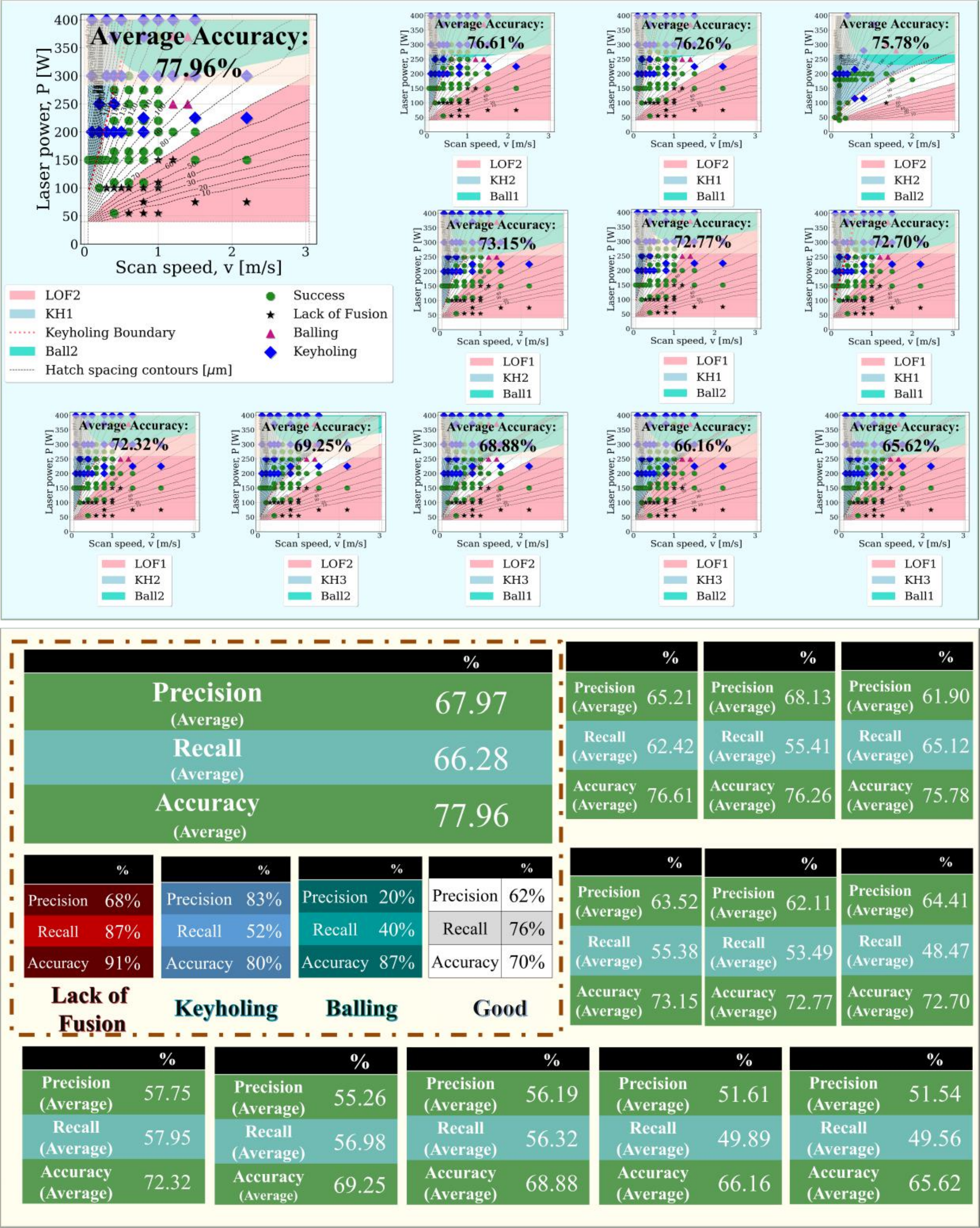}
    \caption{The printability maps produced using different combinations of defect criteria for lack of fusion, balling, and keyholing for Inconel 718.Based on the experimentally observed data and the package prediction in each case, the precision, recall, and accuracy were calculated for four binary classification problems defined based on the presence and absence of each label (i.e., lack of fusion, keyholing, balling, and success or defect-free) in the printability maps and then their averages were listed.  Each map corresponds to the different criteria combination and the maps are placed in order of highest average accuracy to lowest. The values of the precision and recall alongside the accuracy for each map are also displayed. Each map corresponds to the different criteria combination and the maps are placed in order of highest average accuracy to lowest. The values of the precision and recall alongside the accuracy for each map are also displayed. The maximum value for average accuracy was 77.96 \%. The defect labels (i.e., LOF, Ball, and KH) are defined in Table \ref{table:labels}.}
    \label{fig:718 maps}
\end{figure*}

The printability map for Inconel 718 is shown in Figure \ref{fig:718 maps} with a maximum accuracy of 77.96\%. The combination of the criterion that gave the maximum accuracy is Equation \ref{lof2} for lack of fusion, Equation \ref{key1-Gen} for keyholing and Equation \ref{ball1} for balling.

\begin{figure*}
    \centering
    \includegraphics[scale=0.65]{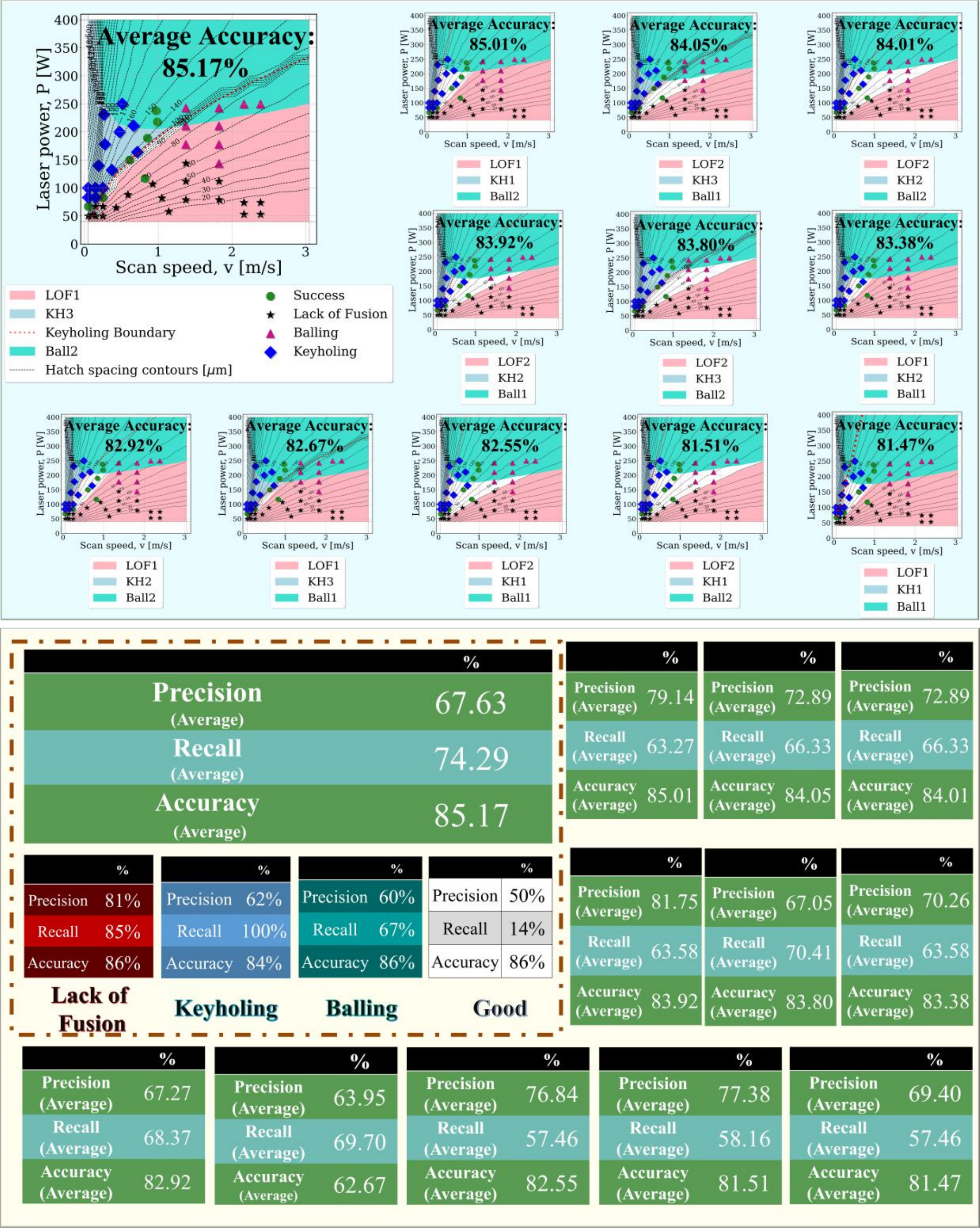}
    \caption{The printability maps produced using different combinations of defect criteria for lack of fusion, balling, and keyholing for AF96.Based on the experimentally observed data and the package prediction in each case, the precision, recall, and accuracy were calculated for four binary classification problems defined based on the presence and absence of each label (i.e., lack of fusion, keyholing, balling, and success or defect-free) in the printability maps and then their averages were listed. Each map corresponds to the different criteria combination and the maps are placed in order of highest average accuracy to lowest. The values of the precision and recall alongside the accuracy for each map are also displayed. The maximum value for average accuracy was 85.17 \%. The defect labels (i.e., LOF, Ball, and KH) are defined in Table \ref{table:labels}.}
    \label{fig:AF96 maps}
\end{figure*}

For AF96, the maximum accuracy, as shown in Figure \ref{fig:AF96 maps}, is 85.17 \% where the minimum accuracy for the material was 81.47 \%. The criteria that give the maximum accuracy are Equation \ref{lof1} for lack of fusion, Equation \ref{Ke} for keyholing and Equation \ref{ball2} for balling. The defect space for the AF96 was thoroughly sampled, similar to 316 SS. 

The Ni-5Nb, the printability map with the maximum accuracy was 74.22 \%. To obtain the maximum accuracy, Equation \ref{lof2} was used for lack of fusion, Equation \ref{Ke} was used for keyholing and Equation \ref{ball2} for balling. The low accuracy can be attributed to the no balling region in the maps as the boundary threshold value set for the current balling region are not optimal. However, to keep the generality of the criteria for the framework demonstration, the criterion was kept as shown in Equation \ref{ball1}.

\begin{figure*}
    \centering
    \includegraphics[scale=0.65]{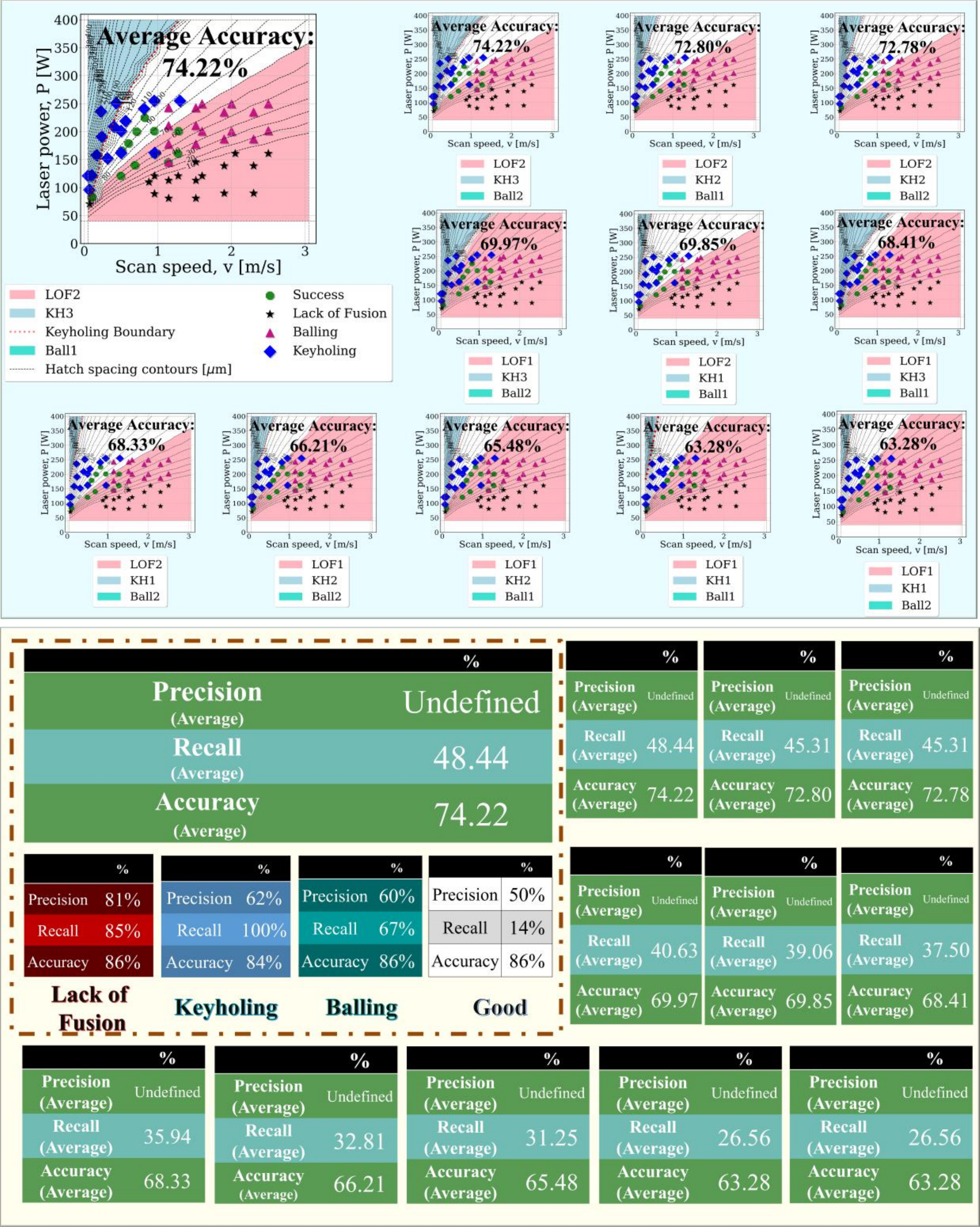}
    \caption{The printability maps produced using different combinations of defect criteria for lack of fusion, balling, and keyholing for Ni-5Nb (weight 5\%).Based on the experimentally observed data and the package prediction in each case, the precision, recall, and accuracy were calculated for four binary classification problems defined based on the presence and absence of each label (i.e., lack of fusion, keyholing, balling, and success or defect-free) in the printability maps and then their averages were listed. Each map corresponds to a different criteria combination and the maps are placed in order of highest average accuracy to lowest. The values of the precision and recall alongside the accuracy for each map are also displayed. For some of the measures, the value could not be calculated due to the lack of data, therefore, the value is marked as "undefined". The maximum value for average accuracy was 74.22 \%. The defect labels (i.e., LOF, Ball, and KH) are defined in Table \ref{table:labels}.}
    \label{fig:Ni5Nb maps}
\end{figure*}

In summary, it can be noted that the prints with experimental data published from literature (316 SS, Ti-6Al-4V, and Inconel 718) contain a higher variance due to different machine parameters associated with each data point; on the other hand, AF96 and Ni-5Nb contain points mixed from published literature and in-house experiments. However, it is also important to note that for alloys such as Inconel 718, Ti-6Al-4V and Ni-5Nb, one set of defects was sparse in the data set. This can be attributed to the fact that most of the published literature will report on experiments that have resulted in successful prints. In comparison, AF96 and 316 SS have a more balanced data set, in which each set of defects was sampled thoroughly. Consequently, the accuracy values reflect these differences. Additionally,  we, note that in general, there was not a single group of criteria that stood out for AM alloys, although there were certainly some criteria that were slightly more commonly observed to correspond to the onset of certain defects. This shows the importance to construct the different printability maps and benchmark the criteria sets for a variety of alloys to have a fuller scope of the processing design space for  unique alloy systems.

\section{Case Study with NiTi-Alloy System}
\label{sec: Case Study}

To further show the effectiveness of the framework, a single alloy system was chosen at different compositions and the corresponding printability maps were constructed and evaluated. 

\subsection{Determining Process Parameters}
\begin{figure*}
    \centering
    \includegraphics[width=0.85\textwidth]{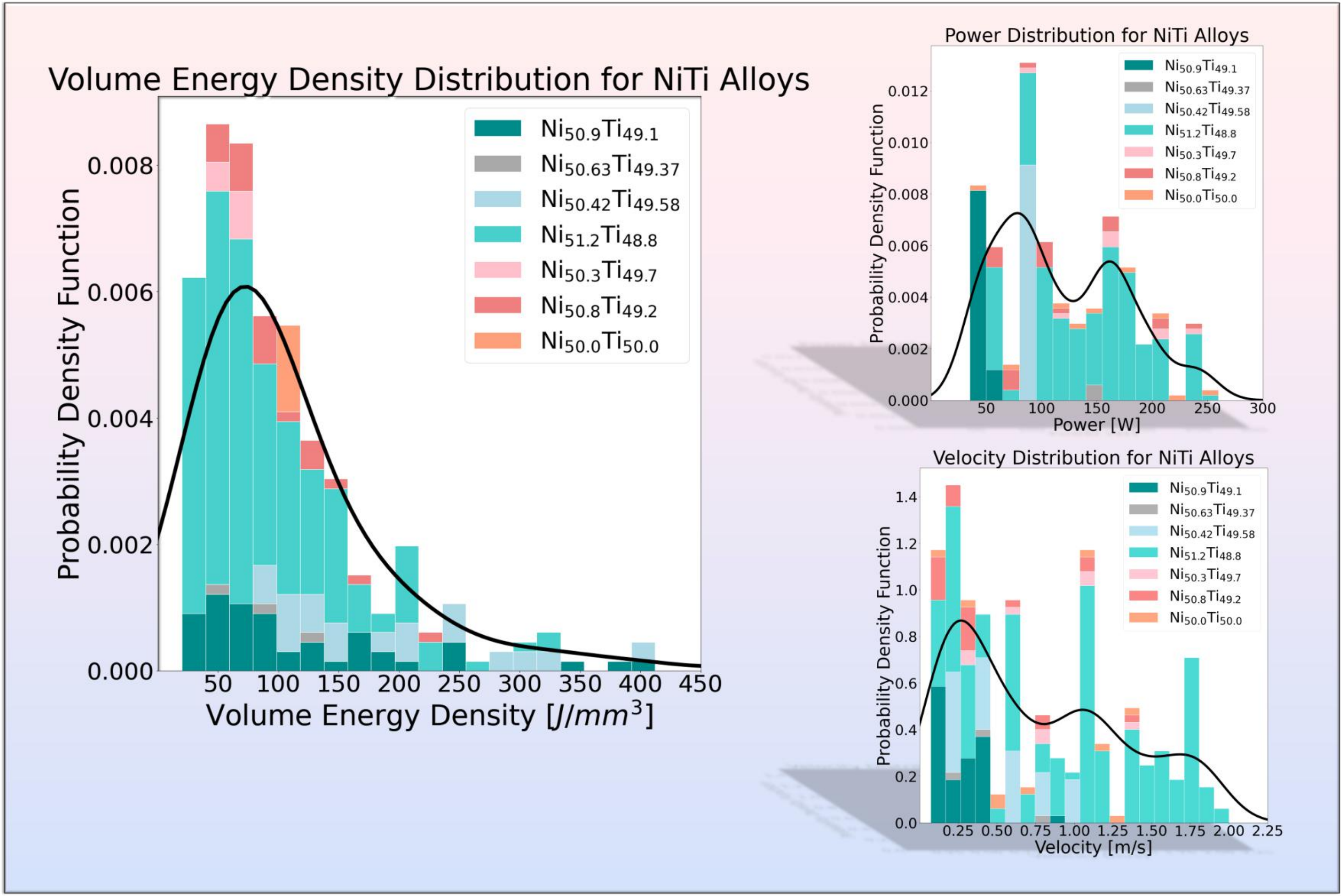}
    \caption{The histogram and kernel density estimate (KDE) plots for laser power, scan speed and volume energy density (VED) that show the distribution of values used to print for a variety of NiTi - based alloys, as reported in literature. The density represents the probability density function of the parameter on the x-axis in the plot.}
    \label{fig:database}
\end{figure*}

To narrow down the \emph{processing} design space for NiTi, experimental data collected for L-PBF AM processes has once again been captured in a database. For NiTi, seven distinct compositions were found in the database, corresponding to 336 data points. The printability map for three of such alloys were fully characterized by the present group in previous published work\cite{xue2021controlling,zhang2022fabrication}. Figure \ref{fig:database} displays the  range of values for laser power and scan speed in the processing design space for an arbitrary NiTi alloy previously printed and/or published in literature. Based on the available data, for the case study of NiTi-based alloy, the processing range considered was 40–300 W and 0.08–2.33 m/s for laser power and scan speed, respectively.

Figure \ref{fig:database} also shows that VED (volumetric energy density, defined as $VED=\frac{P}{v\cdot h \cdot d}$, with $P$ being laser power, $v$ scan speed, $h$ hatch spacing and $d$ layer thickness) peaks within the 50-100 $\frac{J}{mm^3}$, quickly decaying in frequency as soon as VED exceeded $\sim~150~\frac{J}{mm^3}$. In contrast, the distribution of laser power used in the materials was more evenly distributed, ranging from 50 to $\sim$250 W. In the case of scan speed, the conditions used tended to be slightly clustered around relatively low values ($<0.5~m/s$), although there are considerable points in the processing space corresponding to high scan speeds between $1-2~m/s$. It is interesting to note that for some alloys the exploration of the processing space has been exhaustive, in some other cases (e.g. Ni$_{50.9}$Ti$_{49.1}$) the alloys were fabricated within a very narrow process window. 

\begin{figure}
    \centering
    \includegraphics[width=0.4\textwidth]{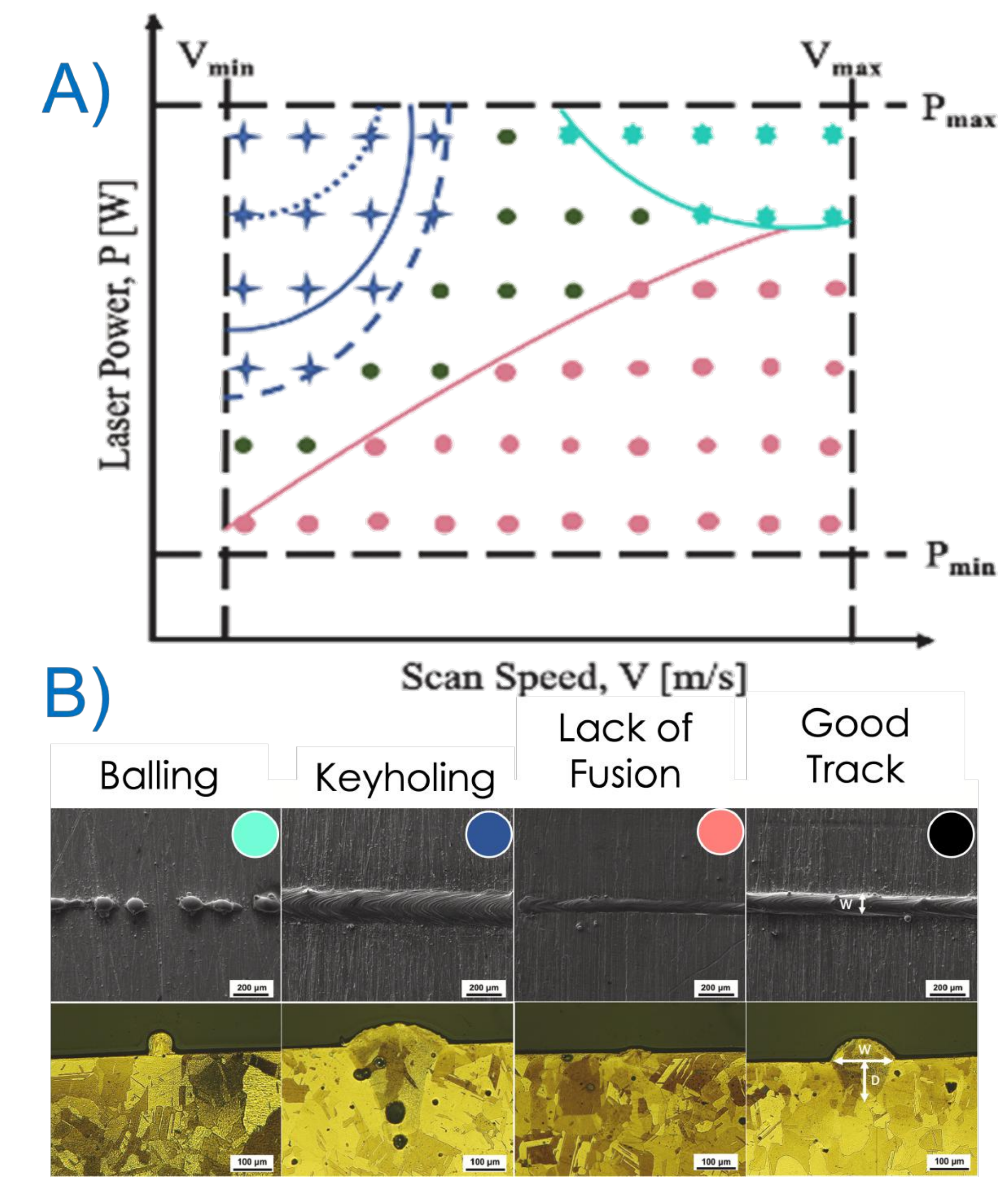}
    \caption{A) Schematic, in process space, of 'ground truth' experiments for the dataset developed in-house. Experiments were designed on a grid containing ~60 points per alloy.B) Morphology and cross-sections for each of the single track experiments were used to classify different points in the process space in terms of the type of defect. Schematic and micrographs were adapted from the original reference~\cite{zhang2021efficient}.}
    \label{fig:tracks}
\end{figure}

After analyzing the data and determining the design space for an arbitrary NiTi alloy using the L-PBF database developed in-house by the co-authors, three different NiTi alloys were chosen to analyze the framework: Ni$_{50.1}$Ti$_{49.9}$ (atomic \%), Ni$_{50.8}$Ti$_{49.2}$ (atomic \%), and  Ni$_{51.2}$Ti$_{48.8}$ (atomic \%). These alloys were fabricated by the present group over the past few years. These alloys were selected on the basis of the quality, breadth and depth of the data used. Other than laser power and speed, machine parameters were kept constant across this data series. Fig.~\ref{fig:tracks} shows a schematic representation of the grid points examined in the in-house printability maps. It is important to note that the 'ground truth' information used to test the models presented here was based on the characterization of the 'top' morphology of single tracks as well as the cross-sections. Each of the tracks was then labeled based on the observed morphology and cross-section. For more details on a typical protocol to characterize the process space used by our group, the reader is referred to a recently published work\cite{zhang2021efficient}. In the present work, physics-based predictions of the printability map using different criteria will be evaluated against this data.

\subsection{Constructing and Evaluating Printability Maps}

\begin{figure*}
    \centering
    \includegraphics[scale=0.65]{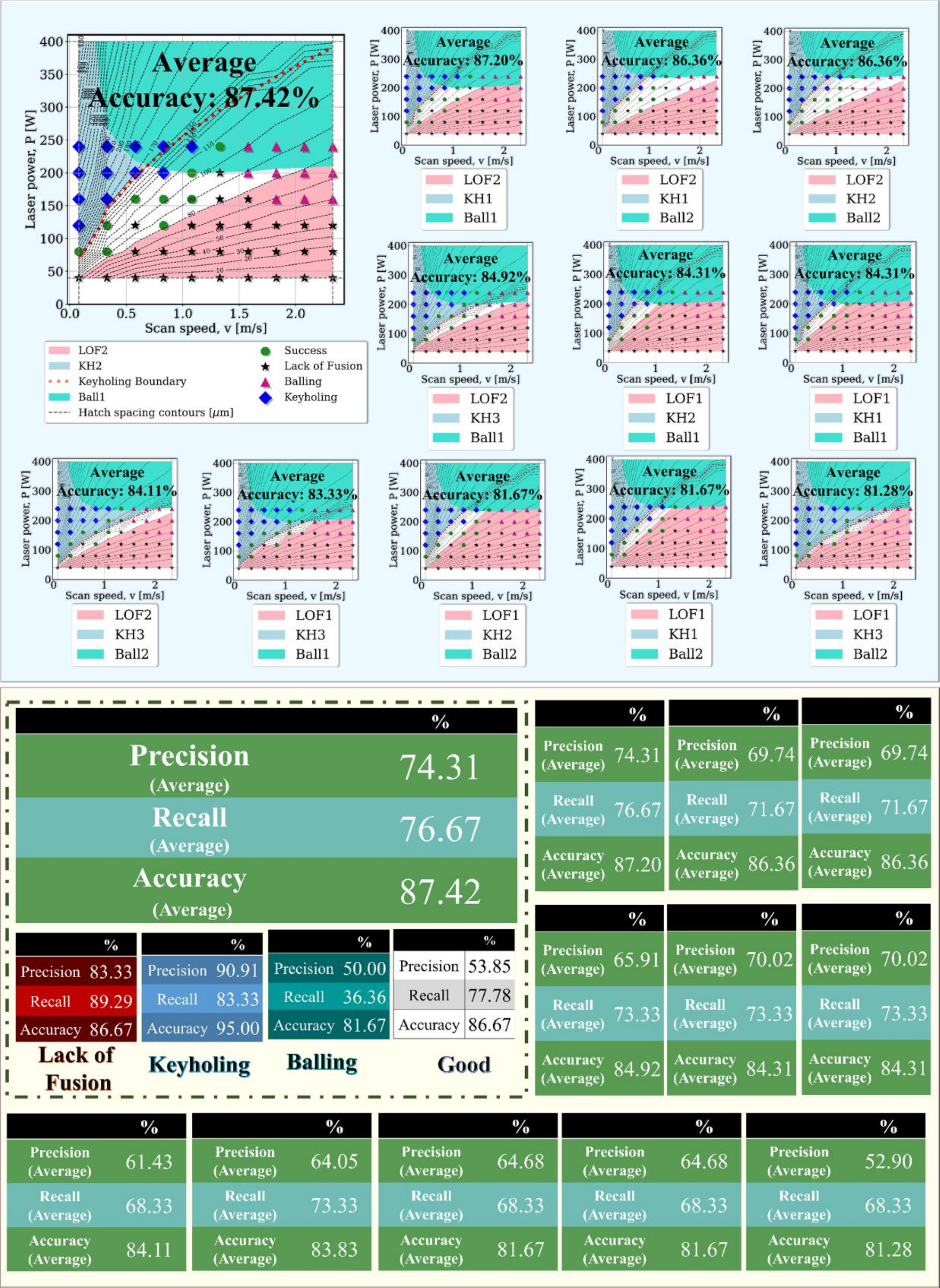}
    \caption{The printability maps produced using different combinations of defect criteria for lack of fusion, balling, and keyholing for  Ni$_{50.8}$Ti$_{49.2}$.Based on the experimentally observed data and the package prediction in each case, the precision, recall, and accuracy were calculated for four binary classification problems defined based on the presence and absence of each label (i.e., lack of fusion, keyholing, balling, and success or defect-free) in the printability maps and then their averages were listed. Each map corresponds to the different criteria combination and the maps are placed in order of highest average accuracy to lowest. The values of the precision and recall alongside the accuracy for each map are also displayed. The maximum value for average accuracy was 87.42 \%. The defect labels (i.e., LOF, Ball, and KH) are defined in Table \ref{table:labels}.}
    \label{fig:50.8 maps}
\end{figure*}

\begin{figure*}
    \centering
    \includegraphics[scale=0.6]{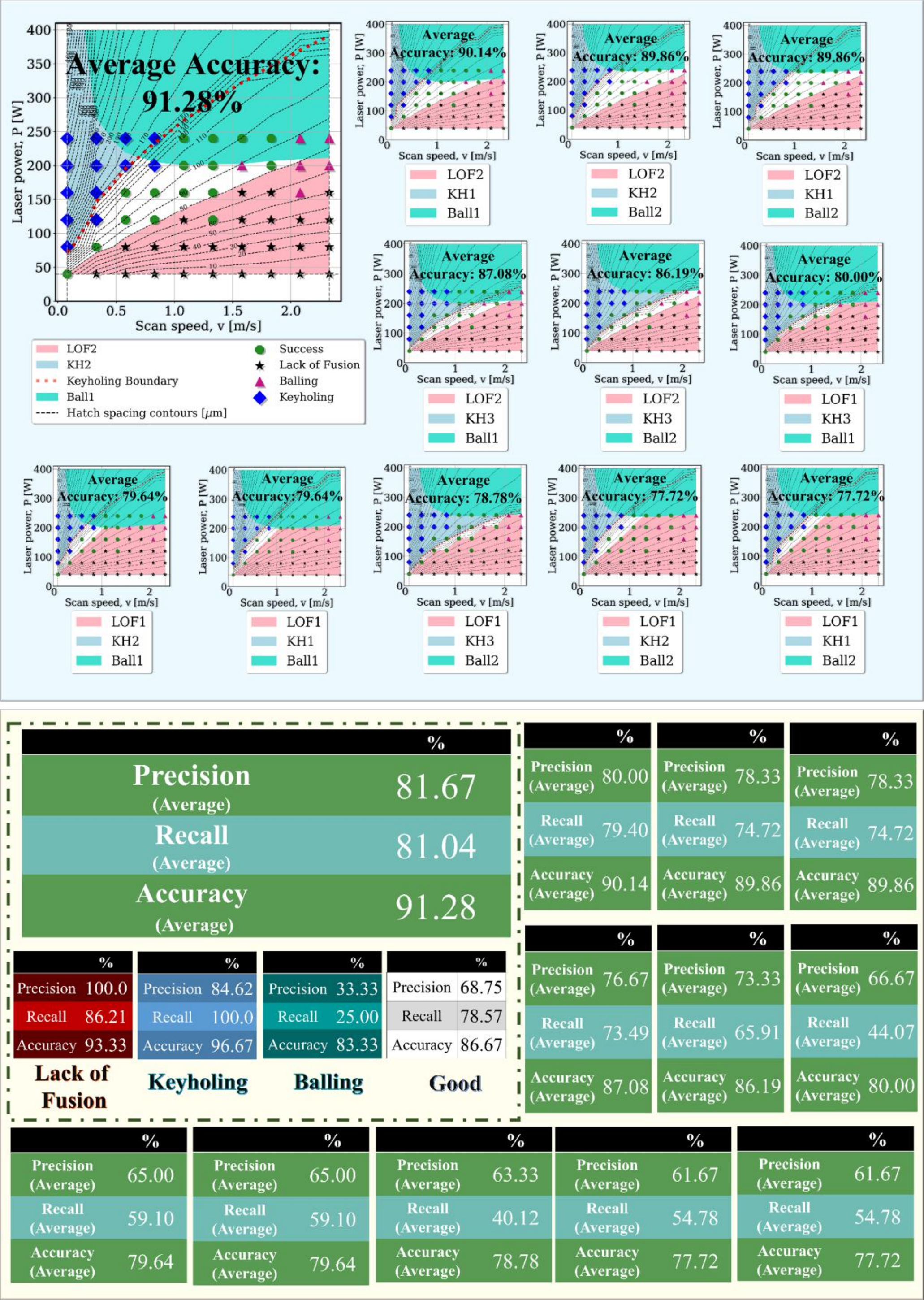}
    \caption{The printability maps produced using different combinations of defect criteria for lack of fusion, balling, and keyholing for  Ni$_{51.2}$Ti$_{48.8}$.Based on the experimentally observed data and the package prediction in each case, the precision, recall, and accuracy were calculated for four binary classification problems defined based on the presence and absence of each label (i.e., lack of fusion, keyholing, balling, and success or defect-free) in the printability maps and then their averages were listed. Each map corresponds to the different criteria combination and the maps are placed in order of highest average accuracy to lowest. The values of the precision and recall alongside the accuracy for each map are also displayed. The maximum value for average accuracy was 91.28 \%. The defect labels (i.e., LOF, Ball, and KH) are defined in Table \ref{table:labels}.}
    \label{fig:51.2 maps}
\end{figure*}

For NiTi-alloy system, three composition were studied to demonstrate the effectiveness of our framework are: Ni$_{50.3}$Ti$_{49.7}$, Ni$_{50.8}$Ti$_{49.2}$, and Ni$_{51.2}$Ti$_{48.8}$. A 10 by 6  design grid of experiments for laser power and scan speed were used to derive the melt pool geometry using the E-T model and the hatch spacing was kept constant at 80 $\mu$m based on experimental data. Printability maps for each combination of the criteria were plotted for all NiTi alloys as shown in Figure \ref{fig:50.3 maps}, Figure \ref{fig:50.8 maps}, and Figure \ref{fig:51.2 maps}.

\begin{figure*}
    \centering
    \includegraphics[width=0.9\textwidth]{Figure10.pdf}
    \caption{The printability maps produced using different combinations of defect criteria for lack of fusion, balling, and keyholing for  Ni$_{50.3}$Ti$_{49.7}$.Based on the experimentally observed data and the prediction in each case, the precision, recall, and accuracy were calculated for four binary classification problems defined based on the presence and absence of each label (i.e., lack of fusion, keyholing, balling, and success or defect-free) in the printability maps and then their averages were listed. Each map corresponds to the different criteria combination and the maps are placed in order of highest average accuracy to lowest. The values of the precision and recall alongside the accuracy for each map are also displayed. The maximum value for average accuracy was 84.39 \%. The defect labels (i.e., LOF, Ball, and KH) are defined in Table \ref{table:labels}.}
    \label{fig:50.3 maps}
\end{figure*}

For Ni$_{50.3}$Ti$_{49.7}$, the maximum accuracy value, as shown in Figure \ref{fig:50.3 maps}, was 84.39\% . The minimum accuracy value for the same system was 81.92\%. The criteria combination that gave the most accurate map for this alloy consisted of Equation \ref{lof2} for lack of fusion, Equation \ref{ball1} for balling, and Equation \ref{key2} for keyholing. For Ni$_{50.8}$Ti$_{49.2}$, the maximum accuracy value was 87.42\%. The criteria combination was the same as the one for Ni$_{50.3}$Ti$_{49.7}$.The minimum accuracy value for this alloy system was 81.28\%. For Ni$_{51.2}$Ti$_{48.8}$, the maximum accuracy value was 91.28\%  where the criteria combination was the same as the one for Ni$_{50.3}$Ti$_{49.7}$ and Ni$_{50.8}$Ti$_{49.2}$. In this case, the minimum accuracy  value was 77.72\%. 

We note that in general, across the three alloys considered, the worst metric corresponded to the recall for balling, followed by the precision for the same type of defect. From the different printability maps, it can be observed that most points labeled as belonging to the balling region were actually at the boundary between balling and the other regions of the processing map (keyholing, lack of fusion, 'good' region). The prediction for this specific region was thus very sensitive to small inaccuracies in the models or criteria used.

If, instead of the maximum average accuracy, the maximum average precision was considered, the same optimal criteria combination would be selected. For Ni$_{50.3}$Ti$_{49.7}$,Ni$_{50.8}$Ti$_{49.2}$, and  Ni$_{51.2}$Ti$_{48.8}$, the maximum precision were 75.40\% , 74.31\%, and 81.67\%, respectively. However, if the maximum average recall is considered, there is a change in the optimal combination criteria for Ni$_{50.3}$Ti$_{49.7}$. As shown in Figure \ref{fig:50.3 maps}, the maximum average recall for Ni$_{50.3}$Ti$_{49.7}$ is 75.00 \%. 

The optimal combination of criteria for the defects was Equation \ref{lof2} for lack of fusion, Equation \ref{ball1} for balling, and Equation \ref{Ke} for keyholing. However, to generalize the optimal combination criteria for NiTi-based alloys, the optimal criteria set for the Ni$_{50.8}$Ti$_{49.2}$ and Ni$_{51.2}$Ti$_{48.8}$ alloys is applied to the Ni$_{50.3}$Ti$_{49.7}$ case and vice versa. These analyses determine which criterion combination is the most optimal one overall. An average recall value of 64.47\% was obtained for the Ni$_{50.3}$Ti$_{49.7}$ case, where the combination of criteria was Equation \ref{lof2}, Equation \ref{ball1}, and Equation \ref{key2}. Comparing this value with the recall value obtained from the consideration of optimal criterion combination results in a difference of 0.89\%. In contrast, the recall values in the case of Ni$_{50.8}$Ti$_{49.2}$ and Ni$_{51.2}$Ti$_{48.8}$, where the criterion set includes Equation \ref{lof2}, Equation \ref{ball1}, and Equation \ref{Ke}, were 64.75\%  and 63.99\% that are 6.94\% and 8.45\% less than the recall values obtained for their optimal criterion set, respectively. Therefore, the optimal combination of criteria for an arbitrary NiTi-based alloy can be concluded to be Equation \ref{lof2}, Equation \ref{ball1}, and Equation \ref{key2}.

\begin{figure*}[htp]
    \centering
         \begin{subfigure}[b]{0.55\textwidth}
         \centering
         \includegraphics[width=\textwidth]{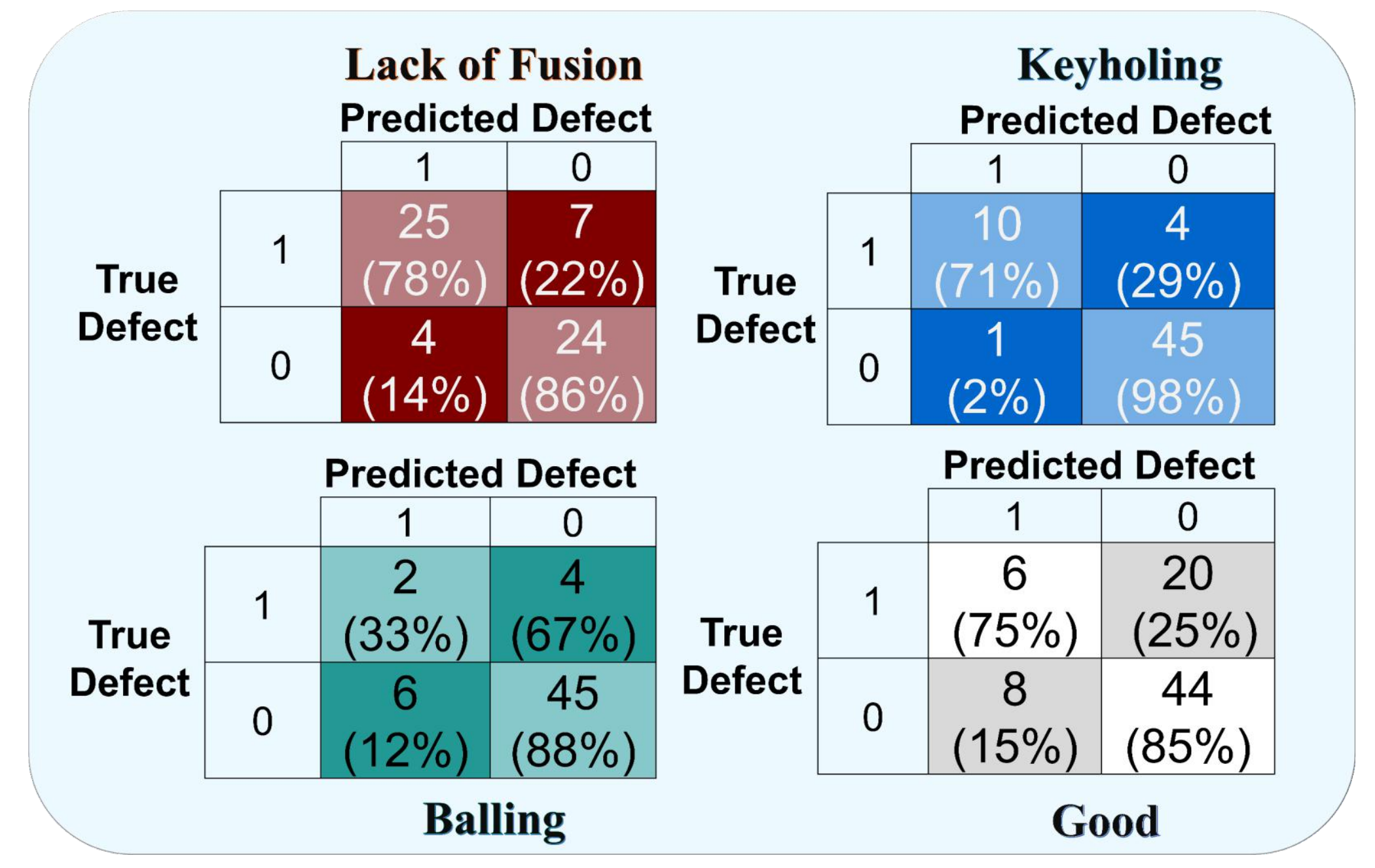}
         \caption{Ni$_{50.3}$Ti$_{49.7}$}
         \label{fig:Conf_50.3}
     \end{subfigure}
     \vspace{0.25mm}
     \begin{subfigure}[b]{0.55\textwidth}
         \centering
         \includegraphics[width=\textwidth]{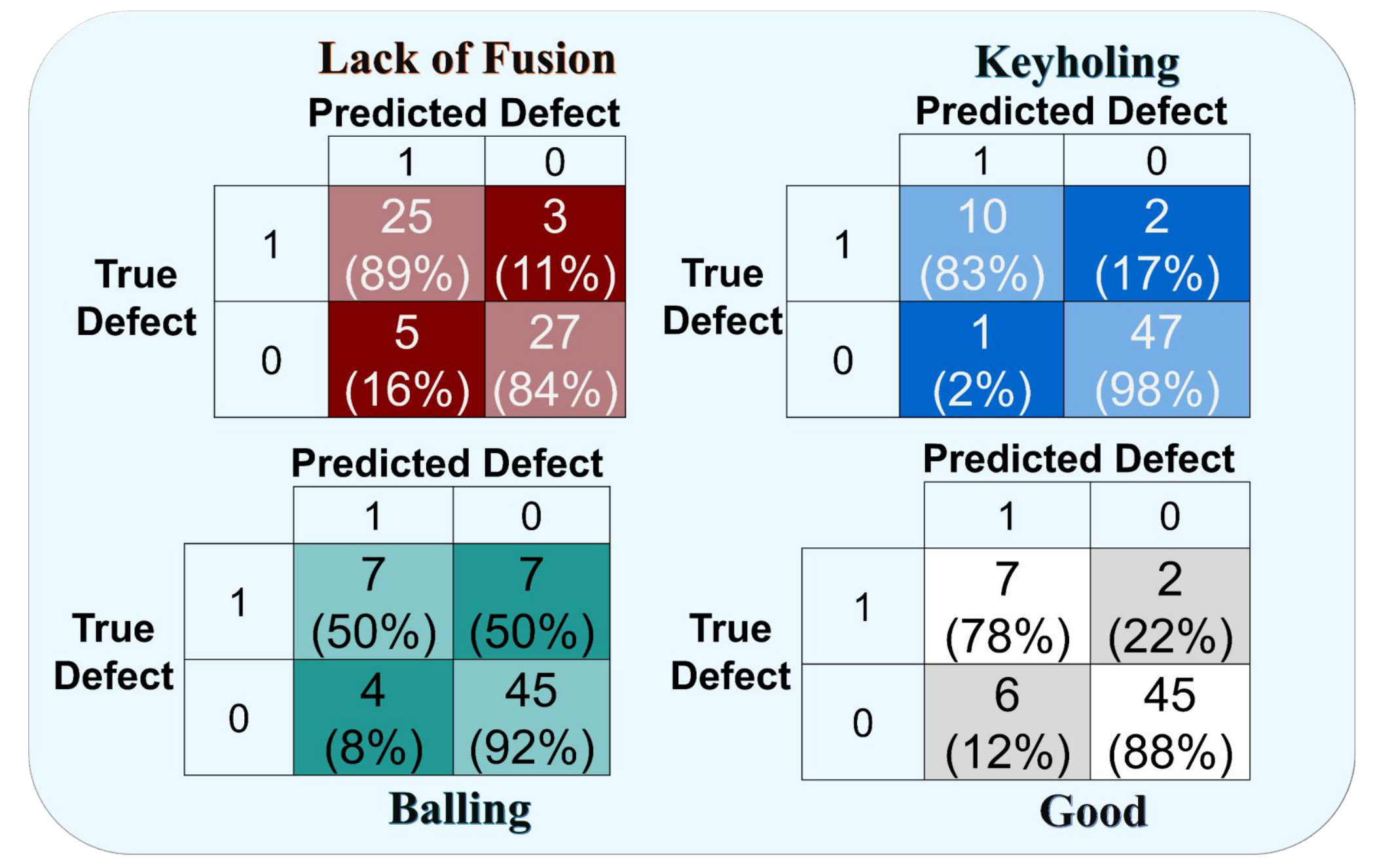}
         \caption{Ni$_{50.8}$Ti$_{49.2}$}
         \label{fig:Conf_50.8}
     \end{subfigure}
     \vspace{0.25mm}
     \begin{subfigure}[b]{0.55\textwidth}
         \centering
         \includegraphics[width=\textwidth]{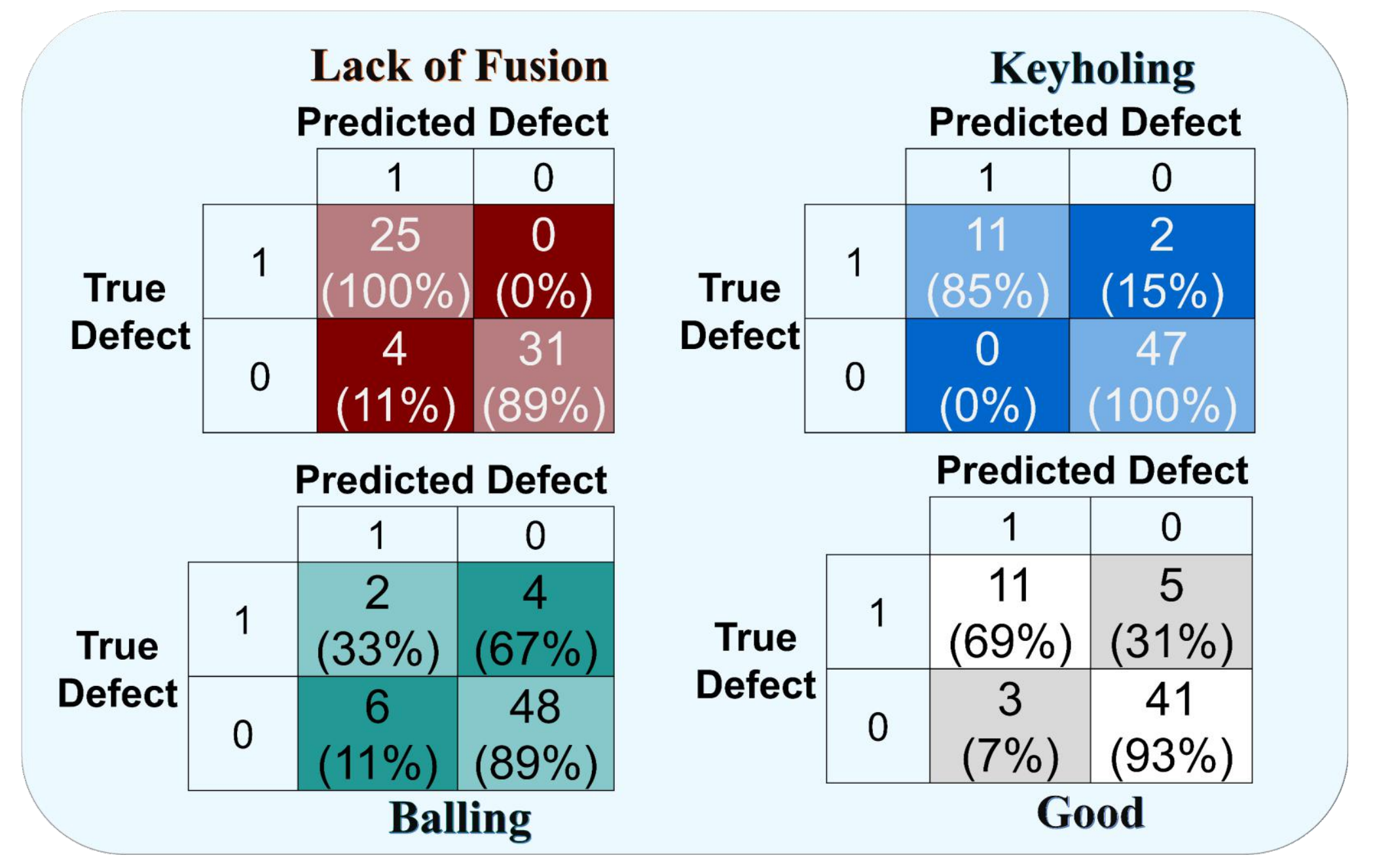}
         \caption{Ni$_{51.2}$Ti$_{48.8}$}
         \label{fig: Conf_51.2}
     \end{subfigure}
        \caption{For the three NiTi alloys, the Confusion matrix is plotted for the optimal combination of criteria, where 1 indicates presence of the respective label in the printability maps and 0 indicates its absence.}
        \label{fig:Conf_Mat}

\end{figure*}

To elaborate how the classification metrics were calculated in this work, the defects' confusion matrices for the optimal combination of criteria for each NiTi alloy were plotted in Figure \ref{fig:Conf_Mat} as examples.

The values for true positive (i.e. when the true defect is 1 and predicted defect is 1), false negative (i.e. when the true defect is 1 and predicted defect is 0), false positive (i.e. when the true defect is 0 and predicted defect is 1), and true negative (i.e. when the true defect is 0 and predicted defect is 0) in the confusion matrices were used to calculate precision, recall, and accuracy for each defect, based on Equations \ref{eq:Pre}, \ref{eq:Rec}, and \ref{eq:Acc}. For instance, in the case of Ni$_{50.3}$Ti$_{49.7}$ alloy, the accuracy for lack of fusion, keyholing, balling, and the 'good' region were 81.67\%,91.67\%, 83.33 \%, and 83.33\%. The accuracy for the binary classifications of defects for the Ni$_{50.8}$Ti$_{49.2}$ alloy were 86.67\%, 95.00\%,81.67\%, and 86.67\% for lack of fusion, keyholing, balling, and the 'good' region. For the Ni$_{51.2}$Ti$_{48.8}$ alloy, the accuracy for lack of fusion, keyholing, balling, and the 'good' region were 93.33\%,96.67\%,83.33\% and 86.67\%. Based on the above accuracy values obtained for the regions in each NiTi printability map, the weighted average accuracy for the optimal criterion set was calculated in Figures \ref{fig:50.3 maps}, \ref{fig:50.8 maps}, and \ref{fig:51.2 maps}. The same calculations were performed for precision and recall metrics.

It should be noted that the proposed criteria set seems to be valid for NiTi-based alloys, however once again, we see that there is a need  to benchmark the proposed criteria with more material systems through an exhaustive comparison with available experimental data. Moreover, there is a need for better defect criteria, specifically when it comes to the keyholing mode, to incorporate the relevant material and processing features missing in the previously used defect criteria. 

While it is recognized that the proposed framework is not perfectly predictive, the resulting printability maps in Section \ref{sec: General AM} and Section \ref{sec: Case Study} tend to offer relatively good agreement with experiments and constitute a valuable first step towards the definition of more precise printability regions. Moreover, in contrast to other approaches, such as those mostly focused on ML-based predictive models, the proposed framework provides interpretable and safe-to-extrapolate predictions that are free of the pathologies that other approaches suffer from, including, for example, discontinuities and other irregularities in the predicted processing maps as well attempt to use more than one criteria set to evaluate the printability maps.

\subsection{Hatch Spacing and Powder Layer Thickness }
\label{sec: hatch-layer}

\begin{figure*}
        \centering
        \includegraphics[width=\textwidth]{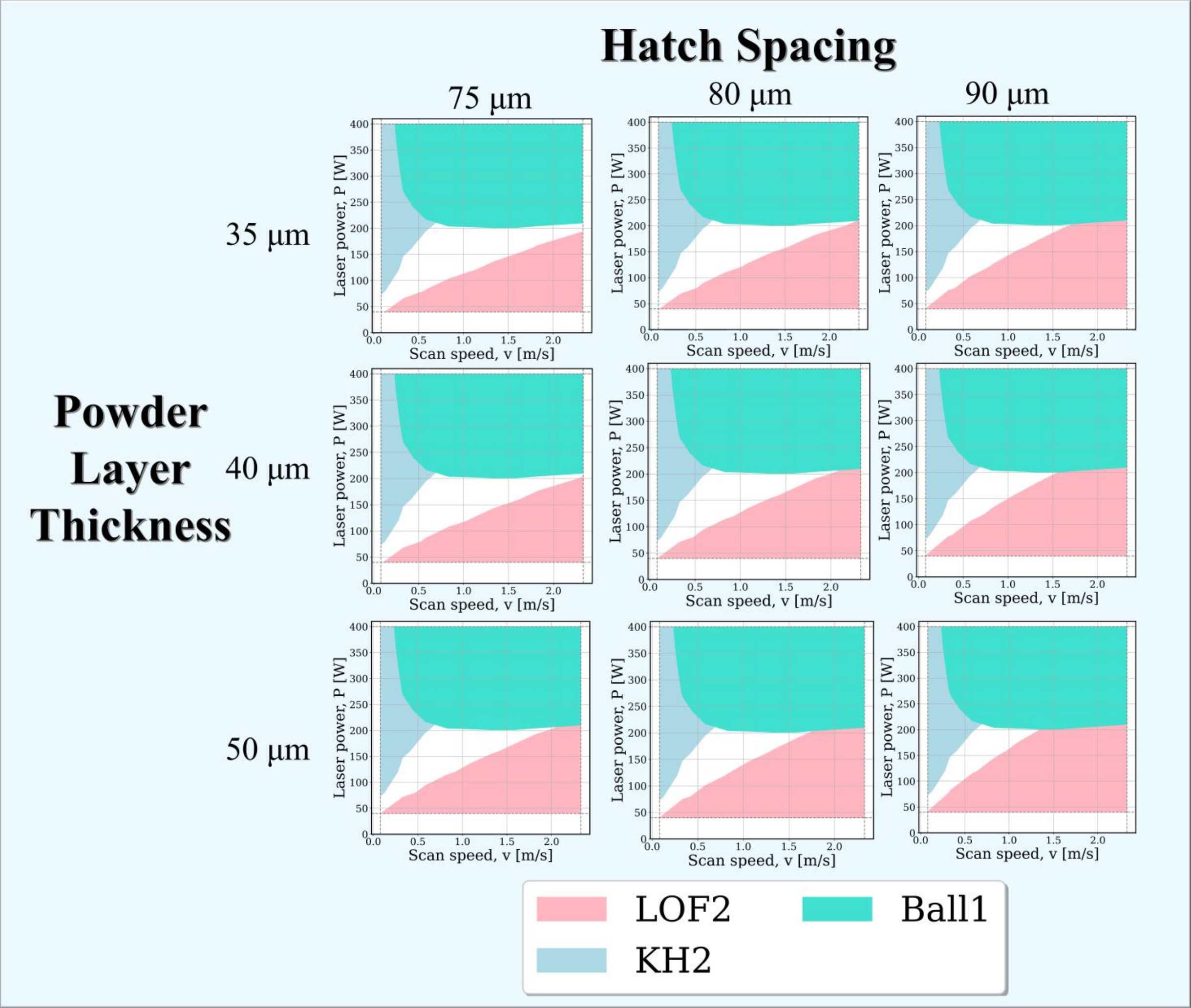}
        \caption{By varying the powder layer thickness and hatch spacing for Ni$_{50.3}$Ti$_{49.7}$, different printability maps can be generated where the defect region (specifically for lack of fusion) changes. The hatch spacing varied from 75 $\mu$m to 90 $\mu$m while the powder layer thickness varied from 35 $\mu$m to 50 $\mu$m. By changing the hatch spacing and powder layer thickness, the defect region defined also changes to create a new printability map.}
        \label{fig:TH}
    \end{figure*}

The printability maps can be further assessed to show the effects of powder layer thickness and hatch spacing. Printability maps were produced for Ni$_{50.3}$Ti$_{49.7}$ with the optimal criteria set.  As shown in \ref{fig:TH}, the hatch spacing varied from 75 $\mu$m to 90 $\mu$m and the powder layer thickness varied from 35 $\mu$m to 40 $\mu$m. The defect-defined region changes with varying hatch spacing and powder layer thickness, namely the lack of fusion region. As the powder layer thickness and/or hatch spacing increases, the area for the lack of fusion region also increases. This can be attributed to the fact that the criterion of Equation \ref{lof2} requires values for hatch spacing and powder layer thickness as inputs. As hatch spacing increases, there is an inadequate overlapping of the melt tracks. Since, the join between adjacent tracks and prior layers is insufficient, lack of fusion induced porosity can occur. Furthermore, with increasing powder layer thickness, more energy input (i.e. higher laser powder) is required to bond the previous layer and/or fully bond the melt pool to the substrate. Therefore, both parameters increase the area for the lack of fusion region in the printability map. 
    
However, there is a need for a formulation and/or criterion that can visualize the effect of hatch spacing and/or powder layer thickness on other defect regions, such as keyholing.

\subsection{Experimentally-calibrated Printability Map vs. Physics-based Printability Map}

In order to better assess the relative value of the proposed predictive framework, it is interesting to compare the resulting printability maps to what would be obtained through a more laborious calibrated printability map. Generally, to construct a calibrated printability map\cite{zhu2021predictive}, an analytical model, such as the E-T model, is implemented and calibrated against experimental data on the melt pool geometry at different process conditions. Then, based on the melt pool geometry from the E-T analytical model, boundaries and regions in the processing space are determined using the experimentally derived criteria for three manufacturing defects---i.e., lack of fusion, balling, and keyholing. 
 
\begin{figure*}
    \centering
    \includegraphics[width=0.85\textwidth]{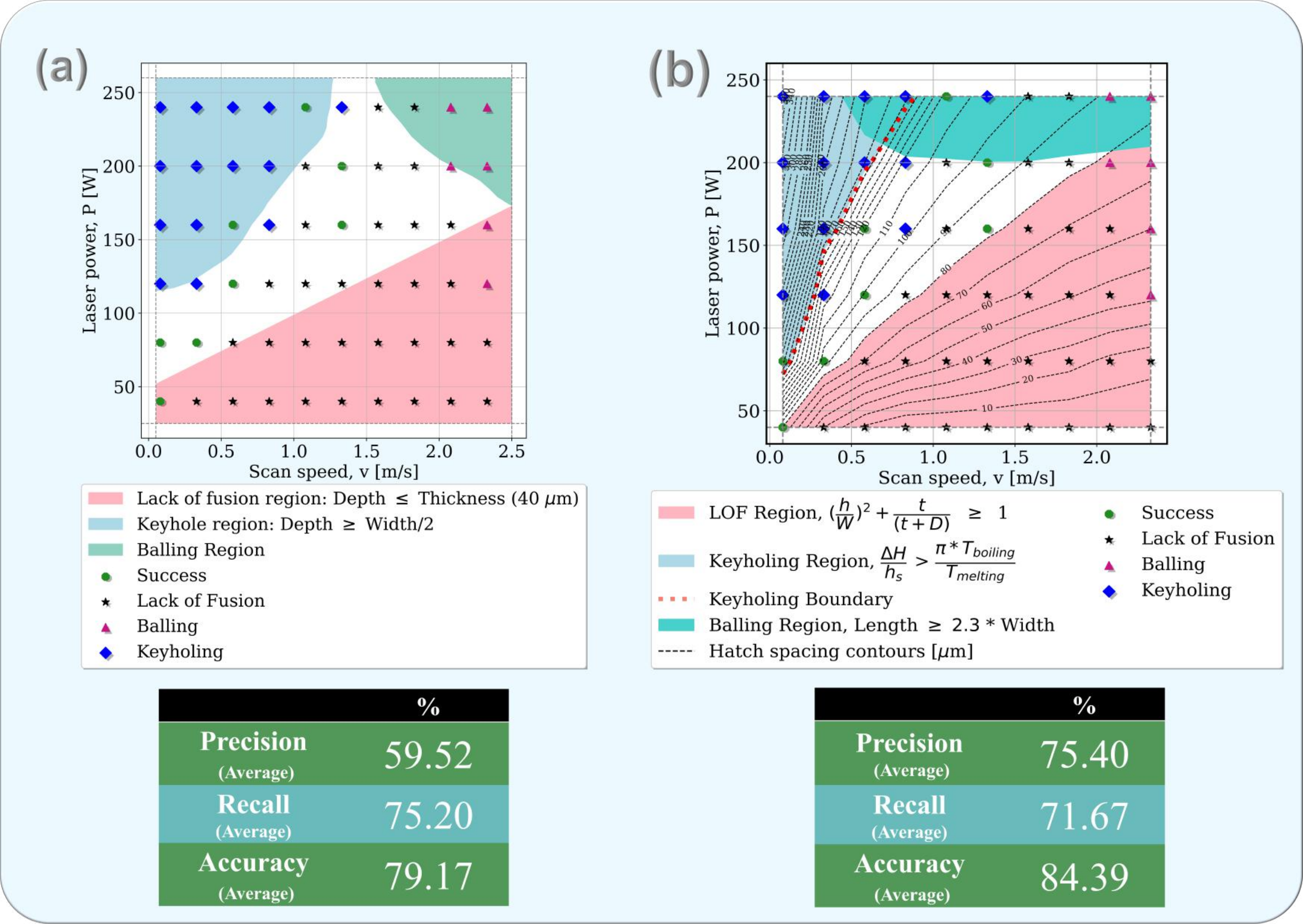}
    \caption{(a) The printability map for Ni$_{50.3}$Ti$_{49.7}$ constructed using a calibrated E-T model, (b) The fully computational printability map for the same alloy generated by the framework presented in this work. For 10 by 10 design grid of laser power and scan speed, the time required to perform experiments and produce the calibrated printability map in part (a) is around 10 days, while it takes just around 2-3 hours to produce the printability map in part (b) using our framework.}
    \label{fig:comp map}
\end{figure*}

For example, to construct the calibrated printability map for Ni$_{50.3}$Ti$_{49.7}$, shown in Figure \ref{fig:comp map}(a), approximately 10 days of work are required due to the need to perform the single track experiments, measure the resulting melt pool dimensions, and then calibrate the E-T model. By contrast, the generation of the corresponding computational-based printability map in Figure \ref{fig:comp map}(b) just takes around 2-3 hours by the framework described in Section \ref{sec:framework}---we note that this computational cost can potentially be reduced to minutes when using ML-based analogues to the E-T model, as will be shown in future work. Moreover, as observed in Figure \ref{fig:comp map}, the values of classification performance metrics calculated for these two maps are very similar. Therefore, the results of our computational framework are almost as accurate as of the previously generated calibrated printability maps, while proving being significantly more resource-efficient.

This provides confidence in the use of this framework for high-throughput analysis of the printability of the entire alloy spaces. We envision future use cases of the proposed computational framework where hundreds or even thousands of alloys can be evaluated for their susceptibility to the onset of AM fabrication defects ahead of their experimental characterization. Such a framework would constitute an important tool within any ICME-based approach to the design of alloys for performance \emph{and} printability. Importantly, and in contrast with  data-only approaches, the boundaries within the printability map can be easily interpreted and directly connected to specific combinations of process conditions and materials properties. This is particularly true when considering ML frameworks based on complex models such as neural networks, as they tend to have poor extrapolation performance over unseen regions of the processing space.

\section{An Example of the Integration Physics-based Printability Maps into ICME Frameworks: Application to Controlled Differential Evaporation of 3D-printed NiTi-based SMAs}
As a tool for cost-effective design, printability maps can be used in combination with material property-process maps to constrain the design space and guide experiments. Previous work by Ranaiefar \etal \cite{ranaiefar2021differential}, for example, established an integrated computational materials engineering (ICME) framework to predict process-structure-property (PSP) relationships and develop martensitic  starting temperature, $M_s$, process maps for L-PBF AM NiTi-based alloys \cite{chen2022processing}. The computational framework exploited the much higher vapor pressure of Ni in Ni-Ti SMAs as the lever to control the composition of the printed part and, thus, its transformation characteristics. In the past, a subset of the present authors have shown that location-dependent control over the feedstock chemistry through differential evaporation can enable 4-D printing of metallic parts\cite{ma2017spatial,atli20194d}.

This ICME framework consists of fast-acting physics-based models and data-driven calibration components. First, a thermal model~\cite{schwalbach2019discrete} enables the simulation of L-PBF AM thermal history and melt pool geometry, for a single track or layer, based on process parameters. Predicted single-track melt pool dimensions and measured single-track experiments are then used to calibrate several parameters: effective thermal conductivity, effective heat capacity, efficiency, and a depth correction factor (used to adjust the keyhole depth). Next, a multi-layer model (MLM) utilizes the output of the calibrated thermal model for a single AM layer to construct the remaining build-layers while simultaneously accounting for any melt pool overlap and corresponding chemistry propagation that occurs. Working in concert with the MLM, the differential evaporation model (DEM) accounts for the loss of alloying elements due to material evaporation during the AM process. In the context of design for NiTi-based alloys, the DEM plays a critical role in the ICME framework and linking PSP relationships. This is due to the highly sensitive Nickel-$M_s$ relationship \cite{frenzel2010influence} and substantial volatility and corresponding propensity of Nickel to evaporate during the AM process \cite{franco2017sensory, ma2017spatial, sam2018tensile}. Consequently, post-process chemistry and properties can differ greatly from the initial powder composition of NiTi-based alloys \cite{chen2022processing, xue2021controlling}, emphasizing the importance of capturing appropriate physics within a model framework to accurately predict PSP relationships and curate process maps expediting AM materials design.

\begin{figure*}
    \centering
    \includegraphics[width=0.85\textwidth]{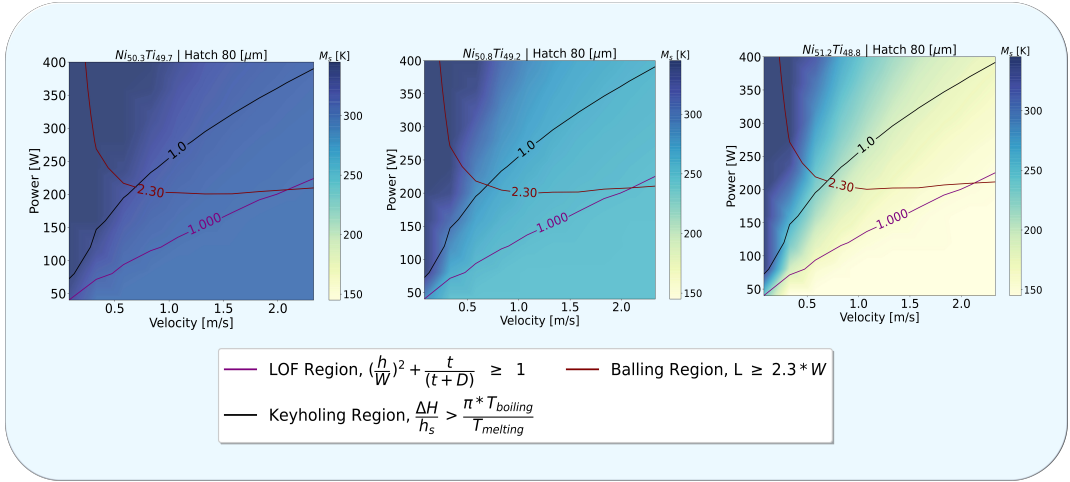}
    \caption{The contour defined for optimal criteria for defects is overlaid on top of $M_s$ process map for Ni$_{50.3}$Ti$_{49.7}$, Ni$_{50.8}$Ti$_{49.2}$, and Ni$_{51.2}$Ti$_{48.8}$ with a constant 80 [$\mu$m] hatch spacing. Analyzing the keyholing region, the combinations of high laser power and low scan speed shows an increase in the $M_s$ values due to increase in Ni evaporation.}
    \label{fig:overlay}
\end{figure*}

Figure \ref{fig:overlay} depicts combined printability and $M_s$ process maps for L-PBF AM Ni$_{50.3}$Ti$_{49.7}$, Ni$_{50.8}$Ti$_{49.2}$, and Ni$_{51.2}$Ti$_{48.8}$ with a constant 80 [$\mu$m] hatch spacing, where lack of fusion, balling, and keyholing regions are constrained by purple, maroon, and black contours, respectively. The contours for lack of fusion, balling, and keyholing were determined based on the optimal combination criteria discussed in Section \ref{sec: Case Study}. Lower energy inputs at regions corresponding to the lack of fusion and balling result in lower evaporation rates and consequently lower $M_s$ temperatures. Corresponding to the keyhole region, combinations of low scan speed and high laser power increased $M_s$ values due to excessive Ni evaporation, peaking at 342 [$^\circ$C] once Ni content falls below the 49.8 [at\%] threshold. Although, relative to Ni$_{50.8}$Ti$_{49.2}$ and Ni$_{51.2}$Ti$_{48.8}$, Ni$_{50.3}$Ti$_{49.7}$ experiences a larger processing region where peak $M_s$ is attainable, this is not fully representative of the NiTi alloy's potential in the context of design. When considering the range of attainable $M_s$ through feedstock chemical control, increasing the initial NiTi alloy composition enables increased tailorability of AM components while maintaining processing parameters within the printable good region. By leveraging process parameters within the printable (acceptable) region and repeated thermal processing, a Ni$_{51.2}$Ti$_{48.8}$ powder could then facilitate the tailored design of location-specific properties for an AM component across the entire range of achievable $M_s$ values.

\section{Conclusion}
 
The present study reports a computational framework to expedite the process of constructing printability maps in metal Additive Manufacturing. The printability maps were defined as the 2D visualization of the processing parameter space (laser power vs. scan speed) that defines regions in which different macroscopic printing defects are prevalent and is an outlook to the processing space of a material under AM conditions.

The framework consists of four steps: calculating material properties, predicting the melt pool geometry using an analytical thermal model, defining a list of criteria for lack of fusion, balling, and keyholing, and constructing a printability map. The material properties are calculated using Thermo-Calc CALPHAD models and a reduced-order model. Using the material properties at the liquidus temperature and defining the laser power-scan speed design grid, the melt pool dimensions---length, width, depth---are calculated using the Eagar-Tsai (E-T) analytical model. Since the E-T model calculates the melt pool at conduction mode, the depth of the melt pool for keyhole mode is underestimated. Therefore, the G-S model was also implemented in the package to estimate the keyhole depth. Furthermore, to define the regions of defect on the map, two criteria for lack of fusion and balling and three criteria for keyholing were applied. Based on the combination of these criteria and the comparison of labels predicted using the package versus experimental observations through the definition of four binary classification problems (where each label in the printability map either exists or not), the optimal combination of criteria to construct printability maps for L-PBF AM alloys. 

The framework was applied to five general AM alloys: 316 Stainless Steel, Inconel 718, Ti-6Al-4V, Ni-5Nb and AF96. It was shown that the predictive capability of the proposed physics-based approach is reasonable across a wide range of alloy systems representative of some of the most commonly investigated alloys in the context of AM. It is noted that the most predictive criteria for the onset of defects is somewhat materials-dependent. This could in part be attributed to the variance in the fabrication factors that was not explicitly taken into account in the analysis presented here. Importantly, these results also show that, to date, there is no unique combination of defect criteria that can be widely adopted in a materials-agnostic manner.

Furthermore, the application of the framework on a single alloy family was shown. In this case, accuracies were significantly higher. This was, in part, because in this case the authors had full knowledge of the experimental parameters and had high quality raw ground truth data to use when evaluating the model+criteria combinations. These last examples also showed how, within a single alloy family, the criteria for the onset of fabrication defects were more generalized across experimental datasets. However, the criterion combinations should be benchmarked with more material systems in the future. There is also room for the improvement of the defect criteria by incorporating more relevant material and processing features.

The accuracy and cost-effectiveness of our computational package were highlighted by simultaneously comparing the data-closeness and the production time of a printability map constructed by our computational framework with its experimentally-calibrated counterpart.

Finally, another layer of complexity was added to the proposed framework by integrating it within an ICME framework for the design of NiTi-based SMAs with specific transformation characteristics, enabled by controlling the differential evaporation of Ni. By introducing an evaporation multi-model framework to predict the final composition and property in the processing space. By overlaying the contours of the defect regions on the process maps, the correlation between defined regions of defect and evaporation rate (tied to the composition and transformation temperature, $M_s$, of the end-use parts), particularly keyholing and excessive evaporation, was shown. Based on such maps, it would be possible to tailor our design by choosing processing parameters such that we are within the printable-good region. In conclusion, the authors have introduced a new fully automated computational framework that will help in expediting the AM product design.
 
\section*{Acknowledgements}
We acknowledge Dr. Bing Zhang and Raiyan Seede for performing several experiments captured in the database used in this work. We also acknowledge David Shoukr for providing the calibrated printability map for an alloy system investigated in this study. Support from the Army Research Office (ARO) under Contract No. W911NF-18-1-0278 is acknowledged. PH and RA also acknowledge the support of NSF, United States, through Grant No. 1849085. MR, BV and SS acknowledge Grant no. NSF-DGE-1545403 (NSF-NRT: Data-Enabled Discovery and Design of Energy Materials, D$^3$EM). The authors would also like to acknowledge the NASA-ESI Program under Grant Number 80NSSC21K0223. High-throughput CALPHAD calculations were carried out in part at the Texas A\&M High-Performance Research Computing (HPRC) Facility.

\bibliographystyle{elsarticle-num}
\bibliography{article.bib}
\end{document}